\documentclass[twocolumn,PRE,showpacs,preprintnumbers,amsmath,amssymb]{revtex4}

\usepackage{graphicx}

\newcommand \bs{\boldsymbol}

\begin{document}

\title{Spatiotemporal patterns in a DC semiconductor-gas-discharge system:\\
Stability analysis and full numerical solutions}

\author{Ismail~R.~Rafatov$^{1,2}$, Danijela D.\ \v{S}ija\v{c}i\'c$^{2,3}$, and Ute Ebert$^{2,4}$}
\email{rafatov@metu.edu.tr,ebert@cwi.nl}
\affiliation{$^1$ Dept.\
Physics, Middle East Technical University, TR-06531 Ankara, Turkey,}
\affiliation{$^2$ CWI, P.O.Box 94079, 1090 GB Amsterdam, The Netherlands,}
\affiliation{$^3$ Dept.\ Applied Earth Sciences, Delft Univ.\ Techn., The Netherlands,}
\affiliation{$^4$ Dept.\ Physics, Eindhoven Univ.\ Techn., The Netherlands.}

\date{\today}

\begin{abstract}
A system very similar to a dielectric barrier discharge, but with a simple stationary DC voltage, can be
realized by sandwiching a gas discharge and a high-ohmic semiconductor layer between two planar electrodes.
In experiments this system forms spatiotemporal and temporal patterns spontaneously, quite similarly to e.g.,
Rayleigh-B\'enard convection. Here it is modeled with a simple discharge model with space charge
effects, and the semiconductor is approximated as a linear conductor. In previous work, this model has
reproduced the phase transition from homogeneous stationary to homogeneous oscillating states
semiquantitatively. In the present work, the formation of spatial patterns is investigated through
linear stability analysis and through numerical simulations of the initial value problem; the methods
agree well. They show the onset of spatiotemporal patterns for high semiconductor resistance.
The parameter dependence of temporal or spatiotemporal pattern formation is discussed in detail.
\end{abstract}

\pacs{05.45.--a, 52.80.--s, 47.54.--r, 02.60.Cb}

\maketitle

\newcommand \be{\begin{equation}}
\newcommand \ee{\end{equation}}
\newcommand \ba{\begin{eqnarray}}
\newcommand \ea{\end{eqnarray}}
\def\nn{\nonumber}
\def\np{\newpage}
\def\bit{\begin{itemize}}
\def\eit{\end{itemize}}
\def\ft{\footnotesize}
\def\vs{\vspace}
\newcommand{\vx}{\vs*{1.0mm}}
\newcommand{\vy}{\vs*{2.5mm}}
\newcommand{\vz}{\vs*{5.0mm}}
\newcommand{\vm}{\vs*{-2.5mm}}
\def\hs{\hspace}
\newcommand{\hx}{\hs*{1.0mm}}
\newcommand{\hy}{\hs*{2.5mm}}
\newcommand{\hz}{\hs*{5.0mm}}






\section{\label{sec:level0}Introduction}

\subsection{Experiments and observations}

Spontaneous pattern formation is a general feature in the natural and technical sciences in systems far from
equilibrium \cite{Cross}. It is a fascinating phenomenon, but can also be detrimental when homogeneity and
stationarity are required in a technical process. Pattern formation occurs frequently in gas discharges, like
in dielectric barrier discharges \cite{Kogel1,Kogel2} that are used, e.g., for ozone production and in plasma
display panels. It is therefore both fundamentally interesting and technically relevant to understand the
mechanisms and conditions of spontaneous symmetry breaking in such systems.

\begin{figure}
\begin{center}
\includegraphics[width=0.5\textwidth]{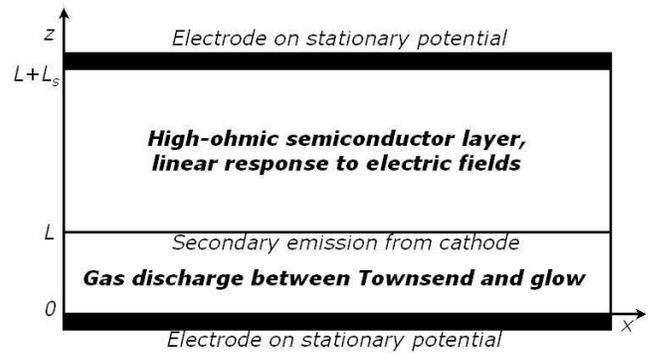} \\
\caption{Scheme of the layers of semiconductor and gas discharge
sandwiched between electrodes with DC voltage.}
\label{fig1}
\end{center}
\end{figure}

A dielectric barrier discharge system consists of a layered structure that in the simplest case consists of a
planar electrode, a dielectric layer, a gas discharge layer, and another planar electrode
\cite{Kogel1,Kogel2}. To the outer electrodes, an AC voltage is applied that forces the system periodically
in time. In the present paper, an even simpler physical system will be analyzed: a system with essentially
the same set-up, but with a DC voltage supply, i.e., a stationary drive. The system is illustrated in
Fig.~\ref{fig1}. To allow current to flow with the DC drive, the dielectric layer is replaced by a high-Ohmic
semiconductor.

Independently of the above technical applications, both the AC and the DC system have attracted much
attention in the pattern formation community in the past two decades, since they are easy to operate, have
convenient length and time scales and a wealth of spontaneously created spatio-temporal patterns. When the
transversal extension of the layers is large enough, experiments show homogeneous stationary and oscillating
modes, and patterns with spatial and spatio-temporal structure like stripes, spots, and spirals
\cite{Willebrand,Gwinn,Str1,Str2,Str3,Str4,StrTh,Astr1,Astr2,Astr3,Astr4,Astr5,Astr6,Astr7,Astr8,Astr9}. The
aspect ratio of a thin layer with wide lateral extension and the observation from above are reminiscent of
Rayleigh-B\'enard convection as the classical pattern forming system in hydrodynamics.

The experiments on DC driven "barrier" discharges in Ref.~\cite{Str1,Str2,Str3,Str4,StrTh} describe not only
phenomena at one particular set of parameters, but in \cite{Str2,StrTh} also explore parameter space and
draw phase diagrams for the transition between different states; therefore we concentrate on those ---  we
are not aware of other experimental investigations of such phase diagrams. The experiments \cite{Str2,StrTh}
are performed on a nitrogen discharge at 40 mbar in a gap of 1 mm. The semiconductor layer consists of
1.5 mm of GaAs. The applied voltages are in the range of 580 V to 740 V. Through photosensitive doping,
the conductivity of the semiconductor layer can be increased by an order of magnitude; the dark conductivity
is $\bar\sigma_s=3.2\times 10^{-8}/(\Omega\, {\rm cm})$. These parameters imply that the product of pressure
and distance $p d$ of the gas discharge is short, but still sufficiently far on the right hand side of the
Paschen curve that the transition from Townsend to glow discharge is subcritical, i.e., that the voltage at
stationary operation drops when the current rises. The resistance of the semiconductor together with the
applied voltage constrain the operation to this transition regime from Townsend to glow discharge.
The system forms stationary states of the discharge that are homogeneous in the transversal direction, as well as
spontaneously oscillating states that still are spatially homogeneous and also oscillating states that show
spatial patterns in the transversal directions as well. In mathematical terms, the last two states emerge
from the homogeneous stationary state through a Hopf-bifurcation (leading to spontaneous oscillations) or
through a Turing-Hopf-bifurcation (leading to spatial and temporal structures). While we investigated the
Hopf-bifurcation in previous work \cite{first,second}, we here include possible structures in the transversal
direction and analyze the occurrence of Turing-, Turing-Hopf- and Hopf-bifurcations.

\subsection{Theoretical approaches and understanding}

Theoretical studies both of the DC and of the AC driven barrier discharge are few; they basically fall into
two classes: $(i)$ simulations or $(ii)$ reduced 2D reaction-diffusion models. $(i)$ Simulations can only be
carried out for particular parameter sets; for these parameters, physical mechanisms of pattern formation can
be identified and visualized. $(ii)$ Reaction-diffusion models \cite{Willebrand,fronts,RGM} in the transversal
2D plane can be investigated analytically, but their derivation from the full 3D physical model depends on
ad hoc approximations whose range of validity beyond the linear regime \cite{fronts} of the Townsend discharge
remains unclear.

In previous \cite{first,second} and in the present work, we have chosen a third line of theoretical
investigation. Namely, $(iii)$ we investigate the linear stability of the homogeneous stationary state and
identify the physical nature of the fastest growing destabilization modes. This allows us to derive complete
phase transition diagrams. This approach is used in many branches of the sciences, but has been little
explored in gas discharges. The stability analysis gives a clue for interesting parameter regimes that can be
further investigated by simulations.

In more physical detail, the state of theoretical understanding and simulations is summarized as follows.

First, for the AC system, the importance of surface charges deposited on the discharge boundaries in each
half-cycle of the voltage drive was identified in \cite{Gwinn} and elaborated in \cite{Ingo1,Ingo2,ChinaAC}.
Simulations use the same fluid models with self generated electric fields as simulations of plasma display
panels \cite{Punset}, most work is performed in two spatial dimensions. Fully three dimensional simulations
have recently been performed in \cite{3DAC}. These simulations are in reasonable agreement with experimental
results, given the uncertainty of the microscopic parameters in the discharge models. Note that even in the
2D system, simulations do not cover many periods of the AC drive, typically not more than 20; therefore, they
have to start from initial conditions fairly close to the final state and cannot follow intermediate
transients in detail.

For the simpler DC driven system, we are not aware of studies of the full discharge model coupled to the
high-Ohmic semiconductor layer, except for a reaction-diffusion model of the system in the two
transversal directions, that is developed, e.g., in \cite{Astr9,fronts}; as said above, this approach does
not appropriately resolve the dynamics in the direction normal to the layers. Of course, surface charges
on the gas-semiconductor interface again have to play a role like in the
AC system, as the electric field can be discontinuous across the interface~\cite{first,second,fronts}. But a
full model needs to account for space charges and ion travel times in the bulk of the gas discharge as well,
cf.~\cite{first,second}.

In our previous work \cite{first,second}, we concentrated on the purely temporal oscillations that occur in a
spatially homogeneous mode, therefore the analysis was restricted to the direction normal to the layers,
assuming homogeneity in the transversal direction. The results presented in \cite{first} showed that a simple
two-component reaction-diffusion approximation for current and voltage in the gas discharge layer is not
sufficient to describe the oscillations, though such a model is suggested through phenomenological analogies
with pattern forming systems like Belousov-Zhabotinski reaction, Rayleigh-B\'enard convection, patterns in
bacterial colonies etc. In \cite{second}, we predicted the phase transition diagram from a homogeneous
stationary to a homogeneous oscillating state. These predictions were in semi-quantitative agreement with the
experiments described in \cite{Str2}.

\subsection{Questions addressed in this paper}

Here we investigate the spontaneous emergence of spatio-temporal patterns in DC driven systems, and predict
in which parameter regimes pattern formation can be expected. While our previous work dealt with the
stationary solutions \cite{Dana,RES} or the dynamics in the single direction normal to the layers
\cite{first,second}, now the transversal direction is included into analysis and simulations as well. The
experiments
described in the paper~\cite{Str2} and thesis~\cite{StrTh} systematically explore the parameter dependence of
pattern formation in the system, and we do the same, but theoretically. The system in~\cite{Str2} never forms
modes that are only spatially structured, but stationary, i.e., it never undergoes a pure Turing transition.
Rather, starting from a homogeneous stationary state, either the homogeneous oscillating state from
\cite{second} is reached through a Hopf-bifurcation, or a spatially structured time-dependent state through a
Turing-Hopf-bifurcation. Furthermore, for high semiconductor resistivity, typically a spatially structured
oscillating state is reached while for low resistivity, the oscillating structures are homogeneous in the
transversal direction. Within the present paper we investigate these transitions first through linear
stability analysis; furthermore interesting parameters regimes are investigated through simulations as
well. We use a fluid model for gas discharge and semiconductor layer coupled to the electrostatic Poisson
equation; in the gas discharge model electrons are adiabatically eliminated to reduce computational costs.

The paper is organized as follows. In Sec.\ \ref{sec:model} we introduce the model, perform dimensional
analysis, and reduce the model by adiabatic elimination of electrons. In Sec.\ \ref{sec:staban}, the problem
of linear stability analysis of the homogeneous stationary state is formulated, the equations are rewritten
and numerical solution strategies are discussed. Sec.\ \ref{StabDisp} contains the results of the stability
analysis, first the qualitative behavior of the dispersion relation as a function of the transversal wave
number $k$, and then predictions on the parameter dependence of the dispersion relations. Sec.\
\ref{sec:compar} presents numerical solutions of the full initial value problem and a comparison with the
stability analysis results; they reveal that both methods can be trusted. Finally, Sec.\ \ref{sec:concl}
contains discussion and conclusions. The numerical details for the solution of the full p.d.e.\ system are
given in the appendix.

\section{\label{sec:model}The Model}

In this section, the simplest model for the full two-dimensional glow discharge--semiconductor system is
introduced. Its schematic structure is shown in Fig.~1. For the gas discharge, it contains electron and ion
drift in the electric field, bulk impact ionization and secondary emission from the cathode as well as space
charge effects. The semiconductor is approximated with a constant conductivity. The same physics was
previously studied, e.g., in \cite{Engel,Raizer} and in our previous papers \cite{Dana,RES,first,second}.
However, in these previous papers, any pattern formation in the transversal direction was excluded and only
the single dimension normal to the layers was resolved. The model then only allows for stationary
solutions \cite{Dana,RES} or temporal oscillations \cite{first,second}. We now study the onset of patterns in
the direction parallel to the layers. If the layers are laterally sufficiently extended, there is rotation
and translation invariance within the plane. Linear perturbations can then be decomposed into Fourier modes.
These Fourier modes can be studied in a two-dimensional setting, i.e., in one direction normal and one
direction parallel to the layers. They are the subject of the paper.

\subsection{\label{subsec:gas_gap}Gas-discharge and semiconductor layers}

The gas-discharge part of the model consists of continuity equations for two charged species, namely,
electrons and positive ions with particle densities $n_{\rm e}$ and $n_{+}$:
\begin{eqnarray}
\partial{_t}n_{\rm e}+\nabla\cdot{\bf \Gamma}_{\rm e} &=& {\it source},
\label{eq:c1}\\
\partial{_t}n_{+}+\nabla\cdot{\bf \Gamma}_{+} &=& {\it source},
\label{eq:c2}
\end{eqnarray}
which are coupled to Poisson's equation for the electric field in electrostatic approximation:
\begin{equation}
\nabla\cdot{\bf E}_g=\frac{e}{\varepsilon_{0}}\left(n_{+}-n_{\rm
e}\right), ~~ {\bf E}_g=-\nabla{\Phi}.
\end{equation}
Here, $\Phi$ is the electric potential, ${\bf E}_g$ is the electric field in the gas discharge, $e$ is the
elementary charge, and $\varepsilon_{0}$ is the dielectric constant. The vector fields ${\bf \Gamma}_{\rm e}$
and ${\bf \Gamma}_{+}$ are the particle current densities, that in simplest approximation are described by
drift only. In general, particle diffusion could be included, however, diffusion is not likely to generate
any new structures, but rather to smoothen out the structures found here. The drift velocities are assumed to
depend linearly on the local electric field with mobilities $\mu_{e}\gg \mu_{+}$:
\begin{eqnarray}
{\bf \Gamma}_{\rm e} = -\mu_{\rm e}n_{\rm e}\,{\bf E}_g,~~~ {\bf \Gamma}_{+} = \mu_{+}n_{+}\,{\bf E}_g,
\end{eqnarray}
hence the electric current in the discharge is
\begin{equation}
\label{eq:Je} {\bf J}_g=e\,\big({\bf \Gamma}_+-{\bf \Gamma}_{\rm e}\big)=e\,\Big(\mu_+n_++\mu_{\rm e}n_{\rm
e}\Big)\,{\bf E}_g.
\end{equation}
Two types of ionization processes are taken into account: the $\alpha$ process of electron impact ionization
in the bulk of the gas, and the $\gamma$ process of electron emission by ion impact onto the cathode. In a
local field approximation, the $\alpha$ process determines the source terms in the continuity equations
(\ref{eq:c1}) and (\ref{eq:c2}):
\begin{equation}
{\it source}=|{\bf \Gamma}_{\rm e}|\,\bar{\alpha}\left(|{\bf E}_g|\right), ~\bar{\alpha}\left(|{\bf
E}_g|\right)=\alpha_{0}\,\alpha\left(\frac{|{\bf E}_g|}{E_0}\right).
\end{equation}
We use the classical Townsend approximation
\begin{equation}
\alpha\left(|{\bf E}|/E_{0}\right)=\exp\left(-E{_0}/|{\bf
E}|\right).
\end{equation}
The gas discharge layer has a thickness $d_g$, where subscripts $g$ or $s$ refer to gas or semiconductor
layer, respectively. The semiconductor layer of thickness $d_s$ is assumed to have a homogeneous and
field-independent conductivity $\bar\sigma_s$ and dielectric constant $\varepsilon_s$:
\begin{eqnarray}
\label{Gsc} {\bf J}_s=\bar\sigma_s{\bf E}_s, ~~~ q=\varepsilon_s\varepsilon_{0}\nabla\cdot{\bf E}_s.
\end{eqnarray}
The space charge density $q$ inside the semiconductor with constant conductivity is assumed to vanish.

\subsection{\label{subsec:bc}Boundary conditions}

In dimensional units, $X$ parametrizes the direction parallel to the layers, and $Z$ the direction normal
to them. The anode of the gas discharge is at $Z=0$, the cathode end of the discharge is at $Z=d_g$, and the
semiconductor extends up to $Z=d_g+d_s$. (Below, in dimensionless units, this corresponds to coordinates
$(x,z)$ and $z=0,~L,~L+L_s$.)

When diffusion is neglected, the ion current and the ion density at the anode vanish. This is described by
the boundary condition on the anode $Z=0$:
\begin{eqnarray}
\label{BC0} {\bf \Gamma}_{+}\left(X,0,t\right)=0 ~~~\Rightarrow~~~ n_{+}\left(X,0,t\right)=0.
\end{eqnarray}
The boundary condition at the cathode, $Z=d_g$, describes the $\gamma$-process of secondary electron
emission:
\begin{eqnarray}
\label{BCdg}
\lefteqn{
\left|{\bf \Gamma}_{\rm e}\left(X,d_g,t\right)\right|
=\gamma\left|{\bf \Gamma}_{+}\left(X,d_g,t\right)\right|} \nn \\
&&\Rightarrow~~~ \mu_{\rm e} n_{\rm e}\left(X,d_g,t\right)= \gamma
\mu_{+}n_{+}\left(X,d_g,t\right).
\end{eqnarray}
Across the boundary between gas layer and semiconductor layer, the electric potential is continuous while the
discontinuity of the normal electric field is determined by the surface charge
\begin{eqnarray}
\Sigma=\Big(\varepsilon_s\varepsilon_{0}{\bf E}_s -\varepsilon_{0}{\bf E}_{g}\Big)\cdot \hat{\bf n},
\label{eq:surf1}
\end{eqnarray}
where $\hat{\bf n}$ is the normal vector on the boundary, directed from the gas toward the semiconductor,
i.e., in the direction of growing $Z$. The change of surface charge in every point $(X,t)$ of the line
$Z=d_{g}$ is determined by the electric current densities in the adjacent gas and semiconductor layers as
\begin{equation}
\partial_{t}\Sigma=\Big({\bf J}_g-{\bf J}_s\Big)\cdot\hat{\bf n},
\label{eq:surf2}
\end{equation}
where ${\bf J}_{g}$ and ${\bf J}_s$ are the current densities at $Z=d_{g}\pm 0$ in gas and semiconductor.
${\bf J}_s$ is given in Eq.\ (\ref{Gsc}), ${\bf J}_g$ on the boundary due to condition (\ref{BCdg}) is \be
\label{eq:surf2a} {\bf J}_{g}\stackrel{Z=d_g}{=} (1+\gamma)\;e\mu_+n_+{\bf E}_{g}. \ee Equations
(\ref{eq:surf1})--(\ref{eq:surf2a}) are summarized as
\begin{eqnarray}
\label{eq:surf3} \Sigma &=& \Big(\varepsilon_s\varepsilon_{0} {\bf E}_s-\varepsilon_{0} {\bf
E}_{g}\Big)\cdot\hat{\bf n}
\\ &=& \Sigma|_{t=0}+\int_{0}^{t}dt~\Big((1+\gamma)en_{+}\mu_{+}
{\bf E}_{g}-\bar{\sigma}_s{\bf E}_s\Big) \cdot\hat{\bf n}.
\nonumber
\end{eqnarray}
This boundary condition is valid in any point $(X,t)$ of the gas-semiconductor boundary $Z=d_g$.

Finally, a DC voltage $U_{t}$ is applied to the gas-semiconductor
system determining the electric potential on the outer boundaries
\begin{eqnarray}
\Phi\left(X,0,t\right)=0,~~~
\Phi\left(X,d_{g}+d_s,t\right)=-U_{t}.
\end{eqnarray}
Here the first potential vanishes due to gauge freedom.

\subsection{\label{subsec:da}Dimensional analysis}

The dimensional analysis is performed essentially as in
\cite{first,second,saarlos,Dana,RES}. However, as it is useful to
measure the time in terms of the ion mobility rather than the
electron mobility, we introduce the intrinsic parameters of the
system as
\begin{eqnarray}
\label{dim}
t_{0}^{(\mu)}=\frac{1}{\alpha_{0}\mu_{\rm +}E_{0}}=\frac{t_0}\mu, ~~~
r_{0}=\frac{1}{\alpha_{0}},~~~
q_{0}=\varepsilon_{0}\alpha_{0}E_{0}.
\end{eqnarray}
Here time immediately is measured in units of $t_{0}^{(\mu)}$,
while in \cite{second}, first the time scale $t_0$ was used.
The intrinsic dimensionless parameters of the gas discharge
are the mobility ratio $\mu$ of electrons and ions and
the length ratio $L$ of discharge gap width and impact ionization length:
\begin{equation}
\mu=\frac{\mu_{+}}{\mu_{e}},~~~L=\frac{d_{g}}{r_{0}}.
\end{equation}
The dimensionless time, coordinates and fields are
\begin{eqnarray}
{\bf r}=\frac{{\bf R}}{r_0},~~~ \tau=\frac{t}{t_0^{(\mu)}},~~~
\sigma({\bf r},\tau)=\frac{e\,n_{\rm e}\left({\bf R},t\right)}{q_{0}}, \nn \\
\rho({\bf r},t)=\frac{e\,n_{+}\left({\bf R},t\right)}{q_{0}},~~~
\bs{\cal{E}}({\bf r},t)=\frac{{\bf E}\left({\bf R},t\right)}{E_{0}}.
\end{eqnarray}
Here the dimensional ${\bf R}$ is expressed by coordinates $(X,Z)$
and the dimensionless ${\bf r}$ by $(x,z)$.

The total applied voltage is rescaled as \be {\cal
U}_{t}=\frac{U_{t}}{E_{0}r_{0}}. \ee The dimensionless parameters of
the semiconductor are conductivity $\sigma_s$ and width $L_s$: \be
\sigma_s=\frac{\bar{\sigma}_s}{\mu_+q_{0}},
~~~L_s=\frac{d_s}{r_{0}}. \ee Note that the dimensionless
conductivity is now also measured on the scale of ion mobility.
Dimensionless capacitance and resistance of the semiconductor and
its characteristic relaxation time are expressed in terms of these
quantities as \be {\cal
R}_s=\frac{L_s}{\sigma_s},~~~C_s=\frac{\varepsilon_s}{L_s},~~~
\tau_s=C_s{\cal R}_s=\frac{\varepsilon_s}{\sigma_s}.
\label{eq:da_RandC} \ee

\subsection{\label{subsec:reduct} Adiabatic elimination of electrons
and final formulation of the problem}

The dynamics of a glow discharge takes place through ion motion
where the ions are much slower than the electrons.
As in \cite{second}, the electrons therefore equilibrate on the
time scale of ion motion and can hence be adiabatically eliminated:
After substituting $s=\sigma/\mu$, the gas discharge part of the
system has the form
\begin{eqnarray} \label{resc1}
\mu\,\partial_\tau s-\nabla\cdot\left(\bs{\cal{E}}s\right)&=&
s |\bs{{\cal E}}|\alpha(|\bs{{\cal E}}|),\\
\partial_{{\tau}}\rho+\nabla\cdot\left(\bs{\cal{E}}\rho\right)&=&
s |\bs{{\cal E}}|\alpha(|\bs{{\cal E}}|),\\
\nabla\cdot{\bs{\mathcal E}} &=& \rho-\mu\,s, ~~~
\bs{{\cal E}} = - \nabla \phi,
\end{eqnarray}
and in the limit of $\mu \rightarrow 0$, it becomes
\begin{eqnarray}
\label{resc2}
-\nabla\cdot\left(\bs{\cal{E}}s\right)&=&
s |\bs{{\cal E}}|\alpha(|\bs{{\cal E}}|),\\
\label{resc22}
\partial_{{\tau}}\rho+\nabla\cdot\left(\bs{\cal{E}}\rho\right)&=&
s |\bs{{\cal E}}|\alpha(|\bs{{\cal E}}|),\\
\label{resc23}
\nabla\cdot{\bs{\cal E}} &=&\rho, ~~~
\bs{{\cal E}} = - \nabla\phi.
\end{eqnarray}
As in \cite{second}, the electric field $\bs{\cal E}$ is now only
influenced by the ion density $\rho$, and not by the much smaller
density of fast electrons (since the electrons are generated in equal
numbers, but leave the system much more rapidly), and the electrons
$s$ follow the ion motion
instantaneously: given the electron density on the cathode $s(x,L,\tau)$
and the field distribution $\bs{\cal E}$ in the gas gap,
the electron density is determined everywhere through (\ref{resc2}).
The boundary conditions (\ref{BC0}) and (\ref{BCdg})
for electrons and ions are
\begin{eqnarray}
\label{2dg05} \rho(x,0,{\tau})=0, ~~~
s(x,L,{\tau})=\gamma \rho(x,L,{\tau}).
\end{eqnarray}

After rescaling, the semiconductor is written as
\begin{eqnarray}
\label{2dg04} \nabla\cdot{\bs{\cal E}}=0,~~~
\bs{\cal E}=-\nabla\phi,~~~
{\bf j}_s= {\sigma_s}\;\bs{{\cal E}},~~~
\end{eqnarray}
and the condition (\ref{eq:surf3}) for the charge
on the semiconductor-gas boundary becomes
\begin{eqnarray}
\label{jump}
&&\frac{\Sigma}{q_0/r_0} = \Big(\varepsilon_s
{\bs{\cal E}}\big|_{z=L^+}-{\bs{\cal E}}\big|_{z=L^-}\Big)
\cdot\hat{\bf n}
\\
\label{jump1}
&&~~= \left.\frac{\Sigma}{q_0/r_0}\right|_{\tau=0} +\int_{0}^{\tau}d\tau
~\Big((1+\gamma)\;\rho{\bs{\cal E}}\big|_{L^-}
-\sigma_s\;{\bs{\cal E}}\big|_{L^+}\Big) \cdot\hat{\bf n}.
\nonumber
\end{eqnarray}
In the perturbation analysis, the differential form of charge conservation
on the boundary is used
\begin{eqnarray}
\label{jumpdiff}
\frac{\partial_{\tau}\Sigma}{q_0/r_0}  =
\left((1+\gamma)\;\rho{\bs{\cal E}}\big|_{L^-} -
\sigma_s\; {\bs{\cal E}}\big|_{L^+}\right)\cdot\hat{\bf n}.
\end{eqnarray}

The total width of the layered structure is $L_z=L+L_s$. On its
outer boundaries $z=0$ and $z=L_z$, the electrodes are on the
electric potential $\phi\left(x,0,\tau\right)=0$ and
$\phi\left(x,L_{\rm z},\tau\right)=-{\cal U}_{t}$, respectively.
Finally, in the numerical solutions of the PDEs, lateral boundaries
at $x=0$ and $x=L_x$ with periodic boundary conditions for $\phi$,
$\rho$, and $s$ are introduced.

\subsection{\label{param} Parameter regime of the experiments}

The parameters are taken as in the experiments~\cite{Str2,StrTh} and in our previous work~\cite{second}.
The discharge is in nitrogen at 40 mbar in a gap of 1 mm. The semiconductor layer consists of 1.5 mm
of GaAs with dielectric constant $\varepsilon_s=13.1$. The applied voltages are in the range of 580--740 V.
Through photosensitive doping, the conductivity of the semiconductor layer could be increased by an
order of magnitude; the dark conductivity was $\bar\sigma_s=3.2\times 10^{-8}/(\Omega\, {\rm cm})$.

For the gas discharge, we used the ion mobility $\mu_{+}=23.33 ~{\rm cm^{2}/(V \, s)}$ and electron
mobility $\mu_{e}=6666.6 ~{\rm cm^{2}/(V \, s)}$, therefore the mobility ratio is
$\mu=\mu_{+}/\mu_{e}=0.0035$. For $\alpha_0=Ap=[27.8\;\mu{\rm m}]^{-1}$ and for
$E_0=Bp=10.3\;{\rm kV/cm}$, we used values from \cite{Raizer}. The secondary emission coefficient
was taken as $\gamma=0.08$. (Comparison with experiment in~\cite{second} as reproduced in Fig.~\ref{fig:bd}
below might suggest $\gamma=0.16$, but that seems unreasonably large.) Therefore the intrinsic scales from
(\ref{dim}) are
\begin{eqnarray}
r_0 = 27.8\;\mu{\rm m},&~ ~&
t_0^{(\mu)} = 11.6\,{\rm ns}, \\
q_0 = 2.04\cdot 10^{12}\,e/{\rm cm^3},&~ ~&
E_0 = 10.3\;{\rm kV/cm},
\nonumber
\end{eqnarray}
the gas gap width of $d=1$ mm corresponds to $L=36$ in dimensionless units, and the semiconductor width
of $d_s=1.5$ mm to a dimensionless value of $L_s=54$.  The dimensionless applied voltages are in the
range of $17.5\leq{\cal U}_t\leq 50$, which correspond to the dimensional range of
$500\,{\rm V}\leq U_t\leq 1425\,{\rm V}$. The dimensionless capacitance of the semiconductor is $C_s=0.243$.
We investigate the conductivity range of $6\cdot10^{-8}/(\Omega\;{\rm cm}) \le \bar\sigma_s\le
6\cdot10^{-7}/(\Omega\;{\rm cm})$ which corresponds to a semiconductor resistance ${\cal R}_s$ of 700
to 7000 in the new dimensionless units (or to $2\cdot 10^5$ to $2\cdot 10^6$ in the units of our previous
papers \cite{first,second}).

For the lowest conductivity of $\bar\sigma_s=3.2\times 10^{-8}/(\Omega\, {\rm cm})$, pattern formation
consistently occurs neither in our analysis nor in the experiment; this case is not discussed further
in this paper.

\subsection{\label{exp} Experiment: between Townsend and glow}

\begin{figure}
\begin{center}
\includegraphics[width=0.4\textwidth]{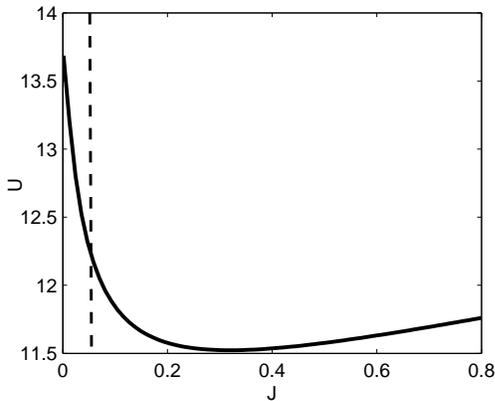} \\
\caption{Solid line: Current voltage characteristics of the gas discharge for $L=36$ and $\gamma=0.08$.
The Townsend voltage ${\cal U}_{\rm Town}=13.7$ has vanishing current $J=0$, the minimal voltage
in the glow regime is ${\cal U}_{\rm glow}=11.5$ at $J\approx0.3$. Dashed line: load line of the semiconductor
for ${\cal U}_t=50$ and ${\cal R}_s=700$. For smaller ${\cal U}_t$ and larger ${\cal R}_s$, the load line
moves closer to the ${\cal U}$ axis, i.e., to the Townsend regime. The intersection of (solid) gas
characteristics and (dashed) load line indicates the stationary solution of the system.}
\label{figIU}
\end{center}
\end{figure}

The parameter regime of the experiments~\cite{Str2,StrTh} as discussed above
can now be placed in the transition regime between Townsend and glow discharge as follows.
We recall \cite{Dana,RES} that the stationary voltage ${\cal U}_{\rm Town}$ of the space charge free
Townsend regime is minimal at a discharge length of $L=e\,\ln\big[(1+\gamma)/\gamma\big]$, which is $L=7.07$
for $\gamma=0.08$ or for example $L=18.8$ for $\gamma=10^{-3}$. (${\cal U}_{\rm Town}(L)$
is known as the Paschen curve.) Continuing with
$\gamma=0.08$ in the remainder of the paper, the transition from Townsend discharge to the space charge
dominated glow discharge is purely subcritical, i.e., the current-voltage characteristic is falling,
if the discharge is longer than $L_{crit}=e^2\,\ln\big[(1+\gamma)/\gamma\big]= 19.23$; this is here
the case. In fact, for the experimental value $L=36$ and $\gamma=0.08$, the Townsend voltage is
${\cal U}_{\rm Town}=13.7$, while the minimum of the voltage in the glow regime is
${\cal U}_{\rm glow}=11.5$, it is reached at a current of $J\approx 0.3$ as shown in figure~\ref{figIU}.
In the experiment the applied voltage does not exceed ${\cal U}_t=50$, and the resistance of
the semiconductor is at least ${\cal R}_s=700$, this situation is indicated as the dashed load line
${\cal U}={\cal U}_t-{\cal R}_s\;J$
in figure~\ref{figIU}. Therefore the current in the stationary homogeneous mode does
not exceed 0.07, i.e., it stays in the transition region between Townsend and glow regime.

\section{\label{sec:staban}Stability analysis: Method}

In this section, the stability analysis of the homogeneous stationary state is set up. While in earlier
work \cite{second}, only temporal oscillations were admitted, here the stability with respect to temporal
and spatial perturbations is analyzed. In particular, linearized equations are derived that define
an eigenvalue problem, and the numerical solution strategy is discussed. It becomes particularly demanding
for large wave numbers $k$. The method developed in this chapter forms the basis for the derivation
of dispersion relations in Chapter~\ref{StabDisp}.

\subsection{\label{subsec:linper}Linear perturbation analysis for transversal Fourier modes:
problem definition}

Here the equations are derived that describe linear perturbations
of the stationary state that is furthermore homogeneous in the transversal
direction, in agreement with the external boundary conditions.
The analysis is set up as in \cite{second}, but now
also transversal perturbations are admitted.
The unperturbed equations are denoted by a subscript 0,
they are 
for $\mu\to0$
\ba
\partial_zj_0&=&-j_0\alpha({\cal E}_0),~~~\mbox{where }j_0=s_0{\cal E}_0,\\
{\cal E}_0\rho_0+j_0&=&J_0,~~~\partial_zJ_0=0,\\
\partial_z{\cal E}_0&=&\rho_0,~~~\partial_z\phi_0=-{\cal E}_0,
\ea
with boundary conditions
\ba
j_0(0)=J_0,~~~&&\phi_0(0)=0,\\
j_0(L)=\frac{1+\gamma}\gamma\;J_0,~~~&&\phi_0(L)=-{\cal U}_0,~~~
{\cal U}_t={\cal U}_0+{\cal R}_sJ_0.  \nn
\ea
Here $J_0$ is the total current and ${\cal U}_t$ is the applied voltage.
For a further discussion, see \cite{Dana,RES,second}.

The first order perturbation theory is denoted by a subscript 1.
As the transversal modes can be decomposed into Fourier modes
with wavenumber $k$ within linear perturbation theory, the ansatz
\begin{eqnarray}
\label{an1}
s(x,z,{\tau}) = s_0(z) + s_1(z)\; e^{ikx+\lambda {\tau}}, \\
\rho(x,z,{\tau}) = \rho_0(z) + \rho_1(z)\; e^{ikx+\lambda {\tau}}, \\
\phi(x,z,{\tau}) = \phi_0(z) +  \phi_1(z)\; e^{ikx+\lambda
{\tau}}.
\end{eqnarray}
is used where the perturbation is supposed to be small.
Insertion of this ansatz into the equations for the gas discharge
(\ref{resc2})--(\ref{resc23}) yields
\begin{eqnarray}
\label{setle}
\partial_z s_1 &=& -\left(\alpha( {\cal E}_0)
+\frac{\partial_z{\cal E}_0}{{\cal E}_0}\right)s_1 -
\frac{s_0}{ {\cal E}_0} \rho_1
\\
&& -\left( \frac{s_0}{{\cal E}_0}\alpha( {\cal E}_0)
+\frac{\partial_z s_0}{{\cal E}_0}+
s_0 \alpha^{\prime}( {\cal E}_0)  \right){\cal E}_{1},
\nn \\
\partial_z \rho_1 &=& \alpha( {\cal E}_0) s_1 -
\left( \frac{\lambda+\rho_0+\partial_z {\cal E}_0}{{\cal E}_0}
\right) \rho_1
\\
&& + \left( \frac{s_0}{ {\cal E}_0}\alpha({\cal E}_0) -
\frac{\partial_z \rho_0}{{\cal E}_0}
+s_0 \alpha^{\prime}({\cal E}_0)\right){\cal E}_{1},
\nn \\
\partial_z {\cal E}_{1}&=& \rho_1 - k^2 \phi_1,
\\
\partial_z \phi_1 &=& -{\cal E}_{1}.
\label{setle4}
\end{eqnarray}
Here ${\cal E}_1$ is the field perturbation in the $z$ direction.
The boundary conditions are:
\begin{equation} \label{8013}
\rho_1(0)=0,~~\phi_1(0)=0~,~~s_1(L) = \gamma \rho_1(L),
\end{equation}
where $z=L$ is the boundary between gas discharge and
semiconductor layer.

In the semiconductor layer, the equation $\Delta \phi = 0$ with the
boundary condition $\phi(L_{z})=-{\cal U}_{t}$ at the position
of the cathode $L_z=L+L_s$ has to be solved. For $\phi_1$,
this means that we have to solve $\Delta\phi_1=0$ with
$\phi_{1}(L_{z})=0$. This problem is solved explicitly for $L
\le z \le L_z$ by
\begin{equation} \label{8014}
\phi_{1}(z) = C_1 \sinh (k(z-L_z)),
\end{equation}
with the arbitrary coefficient $C_1$.
The 'jump' condition (\ref{jump}), (\ref{jumpdiff}) for the electric
field on the semiconductor gas discharge boundary is expressed as
\begin{eqnarray}
\label{jump2a}
&&-C_1 k \cosh (kL_s)\;[\lambda \; \varepsilon_s + \sigma_s]
\nn \\
&&~=\Big[(1+\gamma)(\rho_0 {\cal E}_1
+ {\cal E}_0 \rho_1) + \lambda {\cal E}_1\Big]_L,
\end{eqnarray}
after $\Sigma$ is eliminated.
As the potential (\ref{8014}) is continuous
we get on the boundary $z=L^-$ of the gas discharge region
\begin{equation}
\label{c1} \phi_1(L^+)=\phi_{1}(L^-) = - C_1 \sinh (kL_s).
\end{equation}
Now $C_1$ in (\ref{jump2a}) can be substituted by $\phi_1(L)$
through (\ref{c1}). The result is the second boundary condition at
$z=L$
\begin{equation}
\label{jump2b} \phi_1(L) = \left.\frac{(1+\gamma)(\rho_0 {\cal E}_{1} +
{\cal E}_0 \rho_1) + \lambda {\cal E}_{1}}{\lambda \;
\varepsilon_s+ \sigma_s}\right|_L \;\frac{\tanh (kL_s)}{k}.
\end{equation}
Now the semiconductor is integrated out, and we are left with four
first order ordinary differential equations (\ref{setle})--(\ref{setle4})
and four boundary
conditions (\ref{8013}), (\ref{jump2b}) that together determine
an eigenvalue problem for $\lambda=\lambda(k)$.

\subsection{New fields lead to compacter equations}

It is convenient to write the equations (\ref{setle}) in terms of
the fields $h$ and $g$
\begin{equation}
h = \frac{s_1}{s_0} + \frac{{\cal E}_{1}}{{\cal
E}_0}~~~\mbox{and}~~~ g = {\cal E}_0 \;{\cal E}_{1}
\end{equation}
as in \cite{second}. Furthermore, for non-vanishing wave-numbers $k$, it is convenient to use charge
conservation \be 0=\partial_{{\tau}} \rho +\nabla\cdot\Big(s{\cal E}+\rho{\cal E}\Big)
=\nabla\cdot\Big(\partial_{\tau}{\cal E} +(s+\rho){\cal E}\Big) \ee to eliminate particle densities
completely in favor of the $z$ component of the total current density \be j_1= \lambda {\cal E}_1 +
(s_1+\rho_1) {\cal E}_0 + (s_0+\rho_0) {\cal E}_1. \ee This leads to a compacter form of the system
(\ref{setle})--(\ref{setle4}):
\begin{eqnarray}
\label{2feq}
\partial_z h &=&- \frac{\alpha'}{{\cal E}_0}\;g
- \frac{k^2}{{\cal E}_0}\; \phi_1, \\
\label{2feq2}
\partial_z g &=& - j_0\; h - \frac{\lambda}{{\cal E}_0} \; g +  j_1
- k^2\; {{\cal E}_0}\;\phi_1, \\
\partial_z j_1 &=& -k^2 \; \left(\lambda + \frac{J_0}{{\cal E}_0}\right)
\; \phi_1, \\
\label{2feq4}
\partial_z {\phi}_1 &=& - \frac{1}{{\cal E}_0}\; g.
\end{eqnarray}
The boundary conditions (\ref{8013}) and (\ref{jump2b}) are rewritten as:
\begin{eqnarray}
\label{fbc}
\label{fbc2}
j_1(0) &=& \frac{\lambda}{{\cal E}_0(0)}\; g(0) + J_0\;h(0), \\
\label{fbc0}
\phi_1(0) &=& 0, \\
\label{fbc3}
j_1(L) &=& \frac{\lambda}{{\cal E}_0(L)}\; g(L) +  J_0\;h(L),\\
\label{joj} \label{fbc4}
\phi_1(L) &=& \frac{{\cal R}_s\;j_1(L)}{1+\lambda \;\tau_s}\;
\frac{\tanh (kL_s)}{kL_s}.
\end{eqnarray}
Note that in the last equation, the identity $(\lambda\;\varepsilon_s+\sigma_s)
=(1+\lambda\tau_s)L_s/{\cal R}_s$ was used.
Note further that the limit of $k \rightarrow 0$ of these equations reproduces
the limit of $\mu\to0$ of the analogous 1D equations in \cite{second}.

\subsection{\label{subsec:ni}Formal solution and numerical implementation}

In matrix form, the linearized equations (\ref{2feq})--(\ref{2feq4}) are
\ba
\label{matrix}
&&
\partial_z\;{\bf v}= {\bf M}_{\lambda}\cdot{\bf v},~~~
\mbox{where }{\bf v}(z)=
\left( \begin{array}{c} h \\ g \\ j_1 \\ \phi_1  \end{array} \right)
\\
&& \mbox{and }
{\bf M}_{\lambda} (z) = \left( \begin{array}{cccc}
0 & - \alpha'/{{\cal E}_0}  & 0 & -  k^2/{\cal E}_0  \\
 - j_0 & - \lambda/{{\cal E}_0} &  1 & - k^2 {\cal E}_0 \\
0 & 0 & 0 & -k^2 \left(\lambda+J_0/{\cal E}_0\right) \\
0 & - 1/{\cal E}_0 & 0 & 0 \\
\end{array} \right).
\nn
\ea
The matrix ${\bf M}_{\lambda}(z)$ depends on wavenumber $k$ and
eigenvalue $\lambda$ and total current $J_0$. It depends on $z$ through the functions
${\cal E}_0(z)$, $\alpha({\cal E}_0(z))$, and $j_0(z)$.

The boundary conditions (\ref{fbc2}), (\ref{fbc0}) at $z=0$ mean
that the general solution of the linear equation can be written
as a superposition of two independent solutions of (\ref{matrix})
\ba
\label{eq:super}\label{sol12}
{\bf v}(z) &=& c_1{\bf v}^{(1)}(z)+c_2{\bf v}^{(2)}(z),~~~
\partial_z {\bf v}^{(i)}={\bf M}_{\lambda}\cdot{\bf v}^{(i)},
\nn\\
{\bf v}^{(1)}(0)&=&
\left(\begin{array}{c} 1/J_0 \\ 0 \\ 1 \\ 0 \\ \end{array} \right),~~~
{\bf v}^{(2)}(0) =
\left( \begin{array}{c} 0 \\ {\cal E}_0(0)/\lambda \\ 1 \\ 0 \\
\end{array} \right).
\ea
The solution (\ref{eq:super}) has to obey the boundary conditions
(\ref{fbc3}) and (\ref{joj}) at $z=L$ as well.
Denoting the components of the solutions as
${\bf v}^{(i)}=\big(h^{(i)},g^{(i)},j_1^{(i)},\phi_{1}^{(i)}\big)$
for $i=1,2$, we get
\begin{eqnarray}
\label{eq:system11}
&&c_1\,\left[ j_1^{(1)}-\frac{\lambda}{{\cal E}_0}\; g^{(1)}
-  J_0\;h^{(1)} \right]_{z=L} \nn \\
&&+c_2\, \left[j_1^{(2)}- \frac{\lambda}{{\cal E}_0}\; g^{(2)}
-  J_0\;h^{(2)} \right]_{z=L} = 0, \\
&&c_1\,\left[{\cal R}_s\; j_1^{(1)} - \frac{(1+\lambda \;\tau_s)~k L_s}
{\tanh(k L_s)}~{\phi}_1^{(1)}\right]_{z=L} \nn \\
&&+c_2\,\left[ {\cal R}_s\;j_1^{(2)} - \frac{(1+\lambda\;\tau_s)~k L_s}
{\tanh(k L_s)}~{\phi}_1^{(2)} \right]_{z=L}=0.
\label{eq:system22}
\end{eqnarray}
These equations have nontrivial solutions, if the determinant
\begin{eqnarray}
\label{eq:det}
&&\Delta (z)= \\
&&\left|\begin{array}{cc}
j_1^{(1)}- \frac{\lambda}{{\cal E}_0}\; g^{(1)} -  J_0\;h^{(1)} &
j_1^{(2)}- \frac{\lambda}{{\cal E}_0}\; g^{(2)} -  J_0\;h^{(2)} \\
{\cal R}_s\;j_1^{(1)} - \frac{(1+\lambda \tau_s)~k L_s}{\tanh(k L_s)}~{\phi}_1^{(1)}
& {\cal R}_s\; j_1^{(2)} - \frac{(1+\lambda \tau_s)~k L_s}{\tanh(k L_s)}
~{\phi}_1^{(2)} \\
\end{array} \right|
\nn
\end{eqnarray}
vanishes at $z=L$
\be
\label{eq:detL}
\Delta(z=L)=0.
\ee
Now for a fixed $k$, we start with some initial estimate
for the eigenvalue $\lambda(k)$
and solve equation (\ref{sol12}) numerically for both initial values.
The next estimate for $\lambda$ can be found from condition (\ref{eq:detL})
since it is quadratic in $\lambda$. This process is
iterated until the accuracy is sufficient. We used the condition
$\left|\lambda^{(n+1)}-\lambda^{(n)}\right|
/\left|\lambda^{(n+1)}\right|<10^{-8}$ for the $n$'th iteration step
to finish iterations. For the stability of the iteration process
under-relaxation was used. For the integration of the equations
(\ref{matrix}), we used the classic fourth-order Runge-Kutta method
on a grid with 500 nodes. The majority of investigated solutions
of the present problem are oscillating, therefore the eigenvalues
$\lambda$ are complex, and the eigenfunctions ${\bf v}(z)$ are complex
as well. We have taken this into account by working with complex fields.

After the eigenvalue $\lambda(k)$ is found, the eigenfunction
is determined by inserting the ratio
\begin{eqnarray}
&&\frac{c_2}{c_1}=-\;
\frac{\tanh(kL_s)\;{\cal R}_s\;j_1^{(1)}(L)-kL_s\;
(1+\lambda{\tau}_s)\;\phi^{(1)}_{1}(L)}
{\tanh(kL_s)\;{\cal R}_s\;j_1^{(2)}(L)-kL_s\;
(1+\lambda{\tau}_s)\;\phi^{(2)}_{1}(L)}
\nn \\ &&~
\end{eqnarray}
into Eq.~(\ref{eq:super}).

\subsection{Numerical strategy for large $k$\label{strat}}

The above method gives reliable results for small values of $k$
and has been used to derive the results presented
in Sect.~\ref{ParamDisp}. However, for investigating possible
instabilities for large wave number $k$ as in Sect.~\ref{QualDisp},
e.g., for finding both solution branches for $k>{\cal O}(10^0)$
in Fig.~\ref{fig81}, a better strategy is needed.

There are two points where the integration routine has to be improved
for large $k$.
There is first the fact that the matrix of coefficients is poorly
conditioned. This can be seen by noting that one column is much bigger
than another one. A more precise measure of numerical ill-conditioning
of a matrix is provided by computing the normalized determinant
of the matrix. When the normalized determinant is much smaller than unity,
the matrix is ill-conditioned. The normalized determinant is obtained
by dividing the value of the determinant of the matrix by the product
of the norms of the vectors forming the rows of the matrix.

The second point is the so-called `build-up' error. The difficulty
arises because the solution (\ref{eq:super}) requires combining numbers
which are large compared to the desired solution;
that is ${\bf v}^{(1)}(z)$ and ${\bf v}^{(2)}(z)$ can be up to 3 orders
of magnitude larger than their linear combination,
which is the actual solution. Hence significant digits
are lost through subtraction. This error cannot be avoided
by a more accurate integration unless all computations are carried
out with higher precision. Godunov \cite{Godunov} proposed a method
for avoiding the loss of significance which does not require
multi-precision arithmetics and which is based on keeping
the matrix of base solutions orthogonal at each step of the integration.

A modification of Godunov's method \cite{Conte}, which is
computationally more efficient and  which yields better accuracy,
is implemented in the algorithm used for large $k$.
The main difference to the algorithm described in Section~\ref{subsec:ni}
is that here we examine the base solutions (obtained by any standard
integration method) at each mesh point and when they exceed
certain nonorthogonality criteria we orthonormalize the base solution.
We have to start with initial conditions that are orthogonal to each other
and not only to the boundary conditions:
\ba
{\bf v}^{(1)}(0) &=&
\left( \begin{array}{c}
-1 \\
-1\\
-\frac{\lambda}{{\cal E}_0(0)}-J_0\\
0 \\
\end{array} \right), \nonumber\\
{\bf v}^{(2)}(0)   &=&
\left( \begin{array}{c}
H_0 \\
1 \\
\frac{\lambda}{{\cal E}_0(0)} +J_0 H_0 \\
0 \\
\end{array} \right),
\nonumber
\ea
\be
\mbox{where}~~ H_0 = - \frac{1+\frac{\lambda}{{\cal E}_0(0)}
\left(J_0+\frac{\lambda}{{\cal E}_0(0)}\right) }
{1+J_0 \left(\frac{\lambda}{{\cal E}_0(0)}+ J_0\right)}~.
\ee
Based on the orthogonalization developed in \cite{Conte},
the differential equation (\ref{matrix}) can be solved very accurately
even for large $k$, which allows us to find the eigenvalues
according to the criteria described in the previous section.

Since we must solve the matrix equation
(\ref{matrix}) several times for different values
of $\lambda$, we must insist that the orthonormalization is the same
for all these solutions. In essence we must insist that the determinant
is uniformly scaled in order for the successive approximations
for $\lambda$ to be consistent. Numerically this can be accomplished
by determining the set of orthonormalization points and matrices
for the solution corresponding to the first initial guess for $\lambda$
and thereafter applying the same matrices at the corresponding points
for all successive solutions.

The program is written in C, and the integration method is one-step
simple Runge-Kutta and on the domain L=36 the number of grid points
is varied from n=500 to n=18000 depending on the range of $k$.

\section{Stability analysis: dispersion relations\label{StabDisp}}

\begin{figure}
\begin{center}
\includegraphics[width=0.48\textwidth]{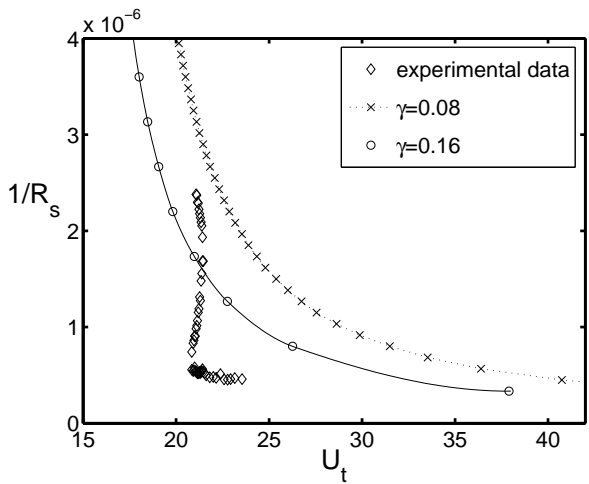} \\
\caption{The Hopf bifurcation lines where the homogeneous stationary state
becomes unstable to homogeneous temporal oscillations; while structure
formation in the transversal direction is excluded.
The bifurcation is drawn as a function of ${\cal U}_t$ and $1/{\rm R}_s$
while all other parameters keep the constant values described in
Sect.~\ref{param}. We recall that ${\cal R}_s$ can be varied by a factor
of 10 by photo-illumination. The oscillations occur above the lines;
shown are two calculated lines for $\gamma=0.08$ and $\gamma=0.16$,
and the experimentally measured line from \cite{Str2}. No parameters have been fitted.
The figure reproduces Fig.~11 from \cite{second}, therefore the old convention for
R$_s={\cal R}_s/\mu$ is used: The upper axis label $1/{\rm R}_s=4\cdot 10^{-6}$
in the plot corresponds to the small ${\cal R}_s=875$.}
\label{fig:bd}
\end{center}
\end{figure}

Having set the mathematical basis for the stability analysis,
now dispersion relations and bifurcation diagrams can be derived
and discussed. In previous work \cite{second}, we have analyzed pure
Hopf transitions where the homogeneous stationary state
becomes unstable to homogeneous oscillations. The bifurcation diagram
for the Hopf transition for the experimental parameters described in
Sections \ref{param} and \ref{exp} is shown in Fig.~\ref{fig:bd},
it is reproduced from Fig.~11 in \cite{second}. Any structures in the
transversal direction were excluded in this earlier work. We now take
this diagram as a basis and investigate which additional spatial
or spatio-temporal instabilities can occur.

\subsection{Hopf or Turing-Hopf bifurcations\label{QualDisp}}

If patterns form spontaneously in the system due to
a linear instability, i.e., in a supercritical bifurcation,
its signature will be found in the dispersion relation $\lambda=\lambda(k)$.
More precisely, there will be a band of Fourier modes
with positive growth rate Re~$\lambda(k)>0$.
If the instability is purely growing or shrinking without oscillations, i.e., if Im~$\lambda(k)=0$,
it is called a Turing instability. On the other hand, if the imaginary
part of the dispersion relation does not vanish (Im~$\lambda(k)\ne0$), the system oscillates.
If the most unstable mode $k^*$ has no spatial structure, $k^*=0$, but only oscillates,
we speak of a Hopf transition, while if $k^*\ne 0$, the transition is called Turing-Hopf.

For three values of ${\cal R}_s$, namely for 700, 1400 and 7000,
the dependence of the dispersion relation $\lambda(k)$ on the applied voltage ${\cal U}_t$
was investigated.

\begin{figure}
  \begin{center}
    \includegraphics[width=0.45\textwidth]{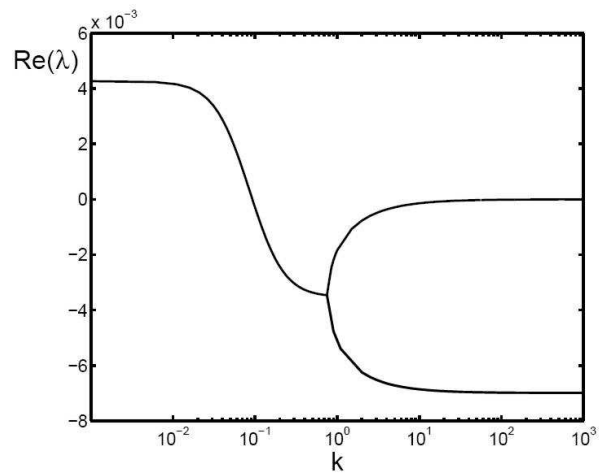} \\
    \caption{Real part of the dispersion relation for ${\cal R}_s=700$
and ${\cal U}_t=23$.}
   \label{disp}\label{fig81}
  \end{center}
\end{figure}

\subsubsection{Qualitative behavior for ${\cal R}_s = 700$ and {\rm 1400}}

For ${\cal R}_s = 700$, a generic shape of the dispersion relation
is presented in Fig.~\ref{disp}. Here a very large range of $k$ values
is shown on a logarithmic scale. The point $k=0$ on the axis
was previously treated in \cite{second}. The dispersion curves $\lambda(k)$
extend continuously from $k=0$ to small $k$: there is a pair of complex
conjugate eigenvalues $\lambda(k)$ that is represented by the single line
for Re$(\lambda)$. For $k$ of order unity, the two complex conjugate solutions
$\lambda(k)$ in Fig.~\ref{disp} merge and form two solutions
with different real part and vanishing imaginary part.

The dynamic behavior is typically dominated by the mode with the largest
positive growth rate Re$(\lambda)$. If the growth rate is negative for all $k$,
the system is dynamically stable. Here the mode $k^*$ with the largest
growth rate has vanishing wave number $k^*=0$ and nonvanishing Im~$\lambda$;
therefore we expect a Hopf bifurcation towards oscillating homogeneous states.
If the upper solution in Fig.~\ref{disp} would develop a positive growth rate
for large $k$, we had found an exponentially growing, purely spatial Turing
instability at short wave lengths, but we haven't found such behavior.

Variation of the applied voltage
${\cal U}_t$ leads to the same qualitative behavior: the
temporally oscillating, but spatially homogeneous mode $k=0$ has
the largest growth rate. Whether this
maximal growth rate is positive or negative, can be read from
Fig.~\ref{fig:bd}. The behavior for ${\cal R}_s = 1400$
is qualitatively the same.

\subsubsection{Qualitative behavior for ${\cal R}_s = 7000$}

\begin{figure}
  \begin{center}
    \includegraphics[width=0.45\textwidth]{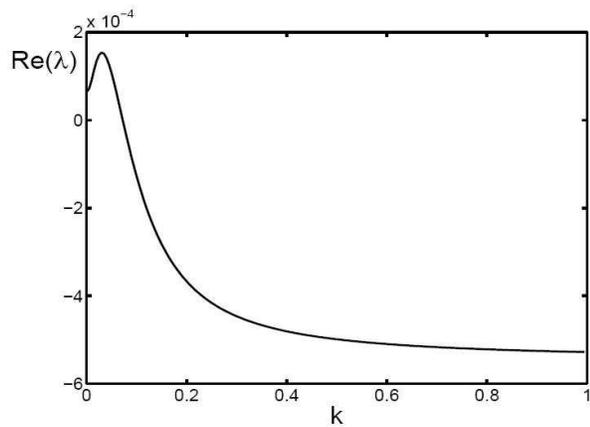} \\
    \caption{Real part of the dispersion relation
for ${\cal R}_s=7000$ and ${\cal U}_t=40$.}
   \label{dispbl}\label{fig82}
  \end{center}
\end{figure}

Further increase of the semiconductor resistivity
to ${\cal R}_s = 7000$ creates a new feature, namely a Turing-Hopf-instability:
the growth rate becomes maximal for some non-vanishing, but very
small value of $k=k^*>0$, as can be seen in Fig.~\ref{dispbl}.

\begin{figure}
  \begin{center}
    \includegraphics[width=0.45\textwidth]{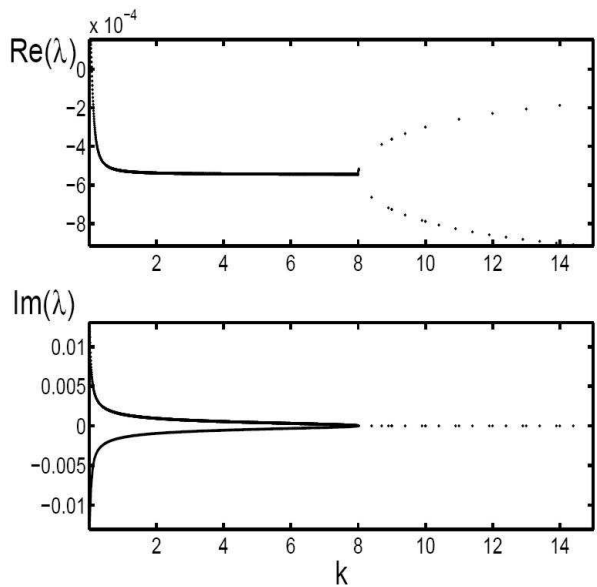} \\
    \caption{Real and imaginary part of the two branches
of the dispersion relation with largest growth rate
for the same parameters as in Fig.~\ref{fig82}.}
   \label{subpl}\label{fig83}
  \end{center}
\end{figure}

In Fig.~\ref{fig83}, real and imaginary part of the two largest
eigenvalues are plotted for a larger range of $k$. Up to $k\approx 8$,
a pair of complex conjugate eigenvalues is found, then two different
branches of purely real eigenvalues emerge, similarly to the behavior
for smaller ${\cal R}_s$ in Fig.~\ref{disp}.

\begin{figure}
  \begin{center}
    \includegraphics[width=0.45\textwidth]{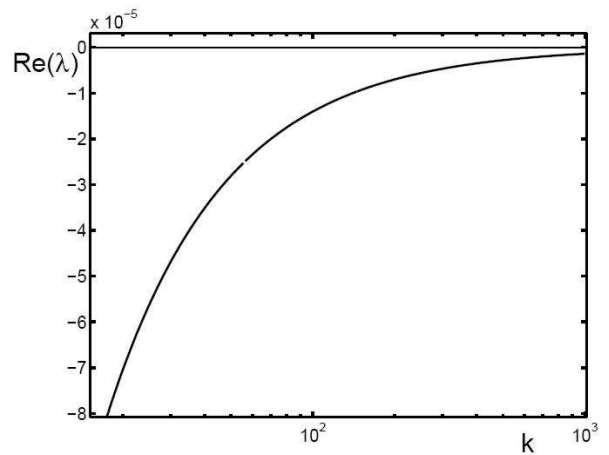} \\
    \caption{Real part of the dispersion relation in the limit
of $k \to \infty$ for the same parameters as in Figs.~\ref{fig82}
and \ref{fig83}.}
   \label{dispblow2}\label{fig84}
  \end{center}
\end{figure}

Finally, Fig.~\ref{fig84} shows that the most unstable branch of
eigenvalues approaches Re~$\lambda(k)=0$ from below for $k\to\infty$,
but does not develop any positive growth rate. Growth rates for large $k$
always stayed negative when exploring the parameters of the system numerically.
If positive growth rates would exist, they would indicate a purely growing spatial
mode with very short wave length.

\subsubsection{Quantitative predictions}

\begin{figure}
\begin{center}
\includegraphics[width=0.4\textwidth]{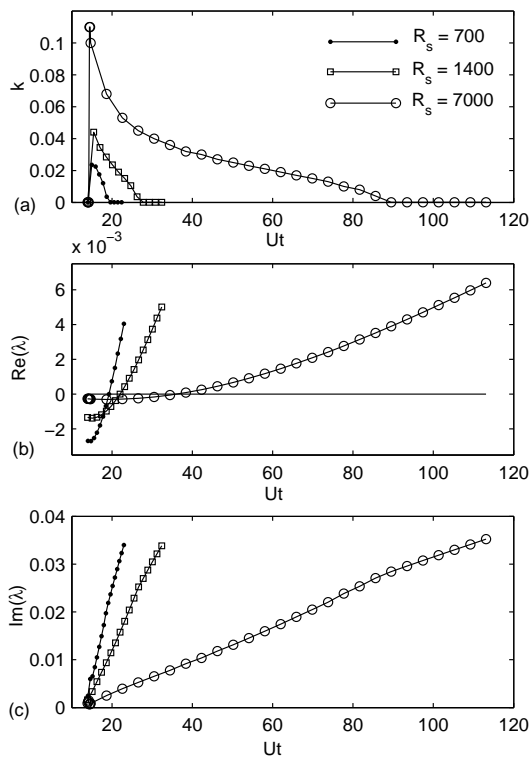} \\
\caption{(a) Most unstable wave number $k^*$, (b) real part and (c)
imaginary part of the corresponding eigenvalue $\lambda(k^*)$ as a
function of the total voltage ${\cal U}_t$ for ${\cal
R}_s=700,~1400,~7000$. The change of ${\cal R}_s$ can be achieved by
photo-illumination.} \label{fig:disprel4}
\end{center}
\end{figure}

The above observations are now quantified in Figure~\ref{fig:disprel4}.
It shows how the most unstable wave number $k^*$ and real and imaginary
part of the corresponding eigenvalue $\lambda(k^*)$
depend on the feeding voltage ${\mathcal U}_{t}$ at different
dimensionless resistances ${\cal R}_s=700$, 1400 and 7000. The
curves begin where the applied voltage ${\cal U}_t$ equals the
Townsend breakdown voltage 13.7, cf.\ Section~\ref{exp}.

Panel (a) shows that for small ${\cal U}_t$ the
most unstable wavelength $k^*$ is always non-vanishing, and that it
decreases for growing ${\cal U}_t$ until it vanishes at some
critical ${\cal U}_t$. Panel (b) shows that the
growth rate Re~$\lambda(k^*)$ of the most unstable mode can stay
negative in the whole range where $k^*>0$. This explains
why the finite wave length instabilities were not seen
above for small ${\cal R}_s$. Panel (c) shows that the most unstable
modes $k^*$ are always oscillating in time.

\begin{figure}
\begin{center}
\includegraphics[width=0.4\textwidth]{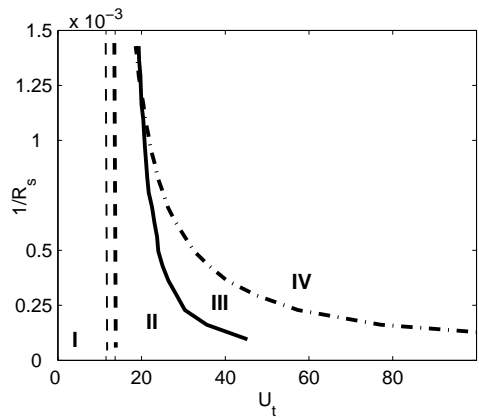} \\
\caption{Calculated bifurcation diagram as a function of voltage ${\cal U}_t$
and conductivity $1/{\cal R}_s$. Dashed lines: minimal voltage ${\cal U}_{\rm glow}=11.5$
(thin) in the glow regime and Townsend voltage ${\cal U}_{\rm Town}=13.7$ (thick)
of the gas discharge according to section~\ref{exp}.
Solid line: Re~$\lambda(k^*)$=0; this line is almost identical with the solid line in
Fig.~\ref{fig:bd} that denotes Re~$\lambda(0)$=0.
Dash-dotted line: $k^*=0$. Region I cannot form
a discharge. In region II, the homogeneous stationary discharge is stable,
in region III, it is Turing-Hopf-unstable, and in region IV, it is Hopf-unstable.
For future investigations, it is interesting to note that all bifurcation lines
become almost straight when plotted as a function of ${\cal R}_s$ rather than
$1/{\cal R}_s$.}
\label{fig:new}
\end{center}
\end{figure}

These results are further summarized in Figure~\ref{fig:new} which is
the counterpart of Fig.~\ref{fig:bd}, but with $1/{\cal R}_s
= 1/({\rm R}_s \mu)$, i.e., with the definition of ${\cal R}_s$ of the present paper.
The solid line in Fig.~\ref{fig:new} is the line
where Re~$\lambda(k^*)$ changes sign. This line is essentially identical
with the solid line in Fig.~\ref{fig:bd} that denotes the sign change
of Re~$\lambda(0)$. The dash-dotted line denotes where the most unstable
wave number $k^*$ vanishes. The dashed lines denote the voltage of
the Townsend and the glow discharge.

The figure shows a calculated bifurcation diagram: In region I,
a discharge cannot form, in region II, the homogeneous stationary discharge is stable,
in region III, it is Turing-Hopf-unstable, and in region IV, it is Hopf-unstable.

\subsubsection{Comparison with experiments\label{CompExp}}

Comparing this calculated bifurcation diagram with the experimental one in \cite{Str2,StrTh},
there is one point in common and one differs.

The common point is that there are no purely spatial patterns without oscillations
in the experiment, in agreement with the theoretical prediction that there is no pure Turing
instability. Furthermore the transition from stationary to oscillating states occurs roughly
in the same parameter regime, cf.~Fig.~\ref{fig:bd}.
The oscillations are due to the following cycle:
$(i)$~the voltage on the gas discharge is above the Townsend value and the ionization
in the discharge increases, $(ii)$~the discharge deposits a surface charge on the
gas-semiconductor interface hence reducing the voltage on the discharge and eventually
extinguishing it, $(iii)$~the high resistance of the semiconductor leads to a long Maxwell
relaxation time of the initial voltage distribution between gas and semiconductor,
after which the cycle repeats.

The difference between experiment and calculation lies in the sequence of temporal
and spatio-temporal patterns. In the experiments, for large conductivity $1/{\cal R}_s$
on increasing ${\cal U}_t$ first purely temporal and then spatio-temporal patterns are formed.
For smaller $1/{\cal R}_s$, the system goes from stationary to oscillating and back
to stationary behavior without loosing homogeneity.
In our calculation, for large $1/{\cal R}_s$, the system directly transits from
homogeneous stationary to homogeneous oscillating, while for small $1/{\cal R}_s$
there is first a range of a Turing-Hopf-instability, and then a pure Hopf-instability
takes over.

On this discrepancy, three remarks are in place.
$(I)$~The nature of the linear instability of the homogeneous stationary system
does not automatically predict the fully developed nonlinear pattern.
$(II)$~Our spatial patterns have wave numbers $k^*$ typically much smaller than 0.1 which
corresponds to wave lengths much larger than 60. Therefore our spatio-temporal
instabilities do not correspond to the dynamic filaments described in \cite{Str2},
but rather to a range of diffuse moving bands. These diffuse moving bands occur before
the blinking filaments discussed in \cite{Str2}; they are only shortly described in the later
Ph.D.~thesis~\cite{StrTh}. We therefore suggest that the moving waves or bands can be identified
with our predicted Turing-Hopf-instability, that creates running waves as well; the experimental data in \cite{StrTh} do not allow
a further comparison with theory and we therefore invite further experimental investigations.
$(III)$~The authors of the experimental papers
have suggested later \cite{Str4,Astr9,Mech} that their experiment is more complex than our simple model
due to nonlinearities in the semiconductor and in gas heating.
Our calculation therefore shows that already this simple model exhibits Hopf- and
Turing-Hopf-instabilities in the same parameter range.

\subsection{Dependence on gap lengths and resistance 
\label{ParamDisp}\label{subsec:dr}}

We now investigate how the dispersion relations change when other system parameters are varied.
While keeping material parameters for gas and semiconductor fixed, the following physical
parameters can be varied: the widths $L_s$ and $L$ of the semiconductor and the gas layer,
the externally applied voltage ${\cal U}_t$ and the semiconductor resistance ${\cal R}_s$
by a factor of 10 through photo-illumination. The dependence on ${\cal U}_t$ was already discussed
above; here the role of the other three parameters is investigated.


\begin{figure}
\begin{center}
\includegraphics[width=0.5\textwidth]{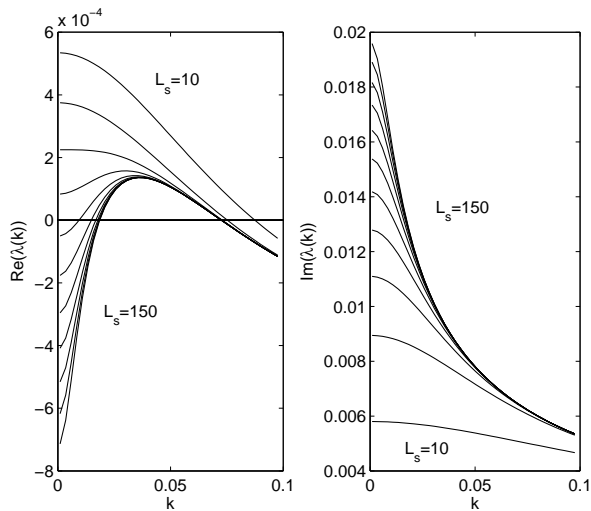} \\
\caption{The influence of the width $L_s$ of the semiconductor layer
on real and imaginary part of the dispersion relation $\lambda(k)$,
for equidistant $L_s$ from the range between $L_s=10$ and 150 at
$J_0=1.32\cdot 10^{-5}$, $L=36$. For $L_s=54$, all parameters are the same
as in Figs.~\ref{fig82}, \ref{fig83} and \ref{fig84}: ${\cal
R}_s=7000$, $C_s=0.243$, and ${\cal U}_{t}=40$.}
\label{fig:disprel1}
\end{center}
\end{figure}

\subsubsection{Dependence on semiconductor width $L_s$}

In Fig.~\ref{fig:disprel1}, we fixed the conductivity at its smallest value
$\sigma_s=54/7000=7.714\cdot 10^{-3}$ and varied $L_s$ from 10 to
150 in equal steps; resistance and capacitance then depend on $L_s$ like
${\cal R}_s\propto L_s$ and $C_s\propto 1/L_s$ according to equation (\ref{eq:da_RandC}) in
section \ref{subsec:da}. In physical units, the width of the
semiconductor layer was changed from 0.28 mm to 4.17 mm.
The length of the gas gap was $L=36$. Rather than the total applied
voltage ${\cal U}_t$, the current $J_0=1.32\cdot 10^{-5}$ was fixed.
For the value $L_s=54$, the corresponding parameter values are
${\cal R}_s=7000$, $C_s=0.243$ and ${\cal U}_{t}=40$ as in
Figs.~\ref{fig82}, \ref{fig83} and \ref{fig84} above.

For small width $L_s$ the system shows a pure Hopf-instability ($k^*{=}0$),
but with increasing $L_s$, the most unstable mode $k^*$ suddenly becomes
nonvanishing, i.e., the system undergoes a first order transition to
spatio-temporal patterns.
Furthermore, the growth rates Re~$\lambda$ decrease and the oscillation frequencies
Im~$\lambda$ increase when $L_s$ increases while the Maxwell relaxation time
$\tau_s={\cal R}_sC_s$ of the semiconductor does not vary.

For the smaller value ${\cal R}_s=700$, a similar behavior is observed:
for the parameters of Fig.~\ref{fig81}, there is still a Hopf bifurcation,
but for increasing $L_s$, the most unstable mode $k^*$ can become positive
as well and a Turing-Hopf-instability occurs.


\subsubsection{Dependence on semiconductor resistance ${\cal R}_s$}

\begin{figure}
\begin{center}
\includegraphics[width=0.5\textwidth]{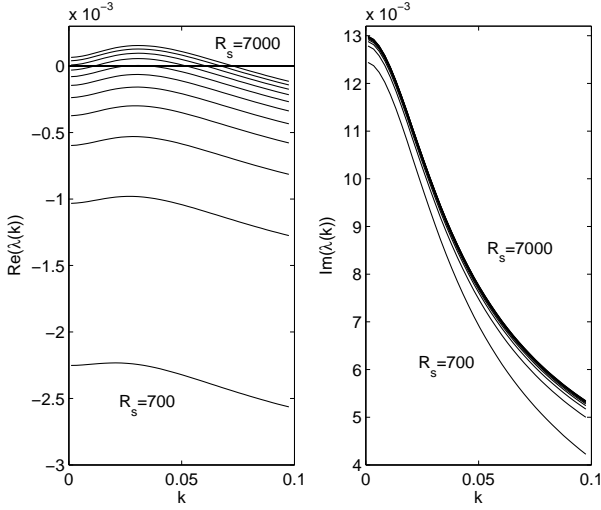} \\
\caption{The influence of the resistance  ${\cal R}_s$ of the
semiconductor layer on real and imaginary part of the dispersion
relation $\lambda(k)$ for equidistant ${\cal R}_s$ in the range
between ${\cal R}_s=700$ and 7000. The current is $J_0=1.32\cdot
10^{-5}$. All other parameters are as in Figs.~\ref{fig82},
\ref{fig83} and \ref{fig84}: $L=36$, $L_s=54$, $C_s=0.243$, and
${\cal U}_t=40$ at ${\cal R}_s=7000$.} \label{fig:disprel11}
\end{center}
\end{figure}

Now the dependence on the conductivity of the semiconductor is tested
that can be varied by illumination by a factor of 10. Accordingly,
in Fig.~\ref{fig:disprel11}, the resistance ${\cal R}_s$ varies in the
interval between 700 and 7000, while the other parameters are chosen
as in the previous section: $L=36$, $L_s=54$, $C_s=0.243$.
The current $J_0=1.32\cdot 10^{-5}$ is fixed; this value corresponds
to a voltage of ${\cal U}_t=40$ for ${\cal R}_s=7000$. Therefore,
the upper line in Fig.~\ref{fig:disprel11} is the same as the one in
Figs.~\ref{fig82}, \ref{fig83} and \ref{fig84}.

For all values of the resistivity, the wave number $k^*$ where the
growth rate Re~$\lambda$ is maximal, stays positive ($k^*>0$), but it drops
below zero for smaller ${\cal R}_s$ making the homogeneous stationary state
stable. Furthermore, for the smallest investigated ${\cal R}_s$, namely
${\cal R}_s=700$, and fixed $J_0$, the voltage is ${\cal U}_t=16.25$,
but when ${\cal U}_t=23$ the growth rate is positive for the same $k$ range
and has maximum for $k=0$ as can be seen in Fig.~\ref{fig81}.


\begin{figure}
\begin{center}
\includegraphics[width=0.5\textwidth]{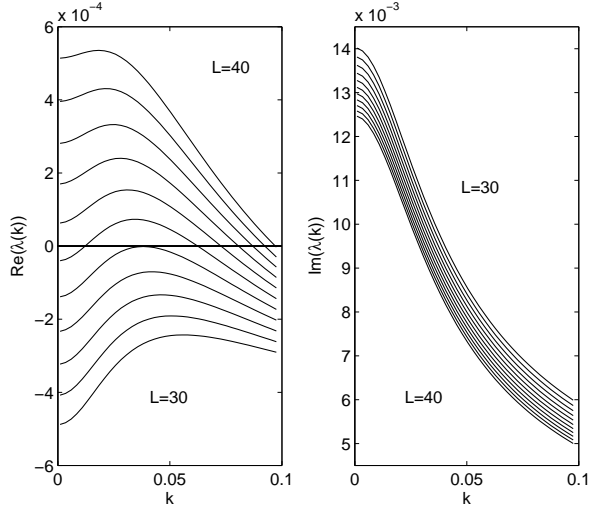} \\
\caption{The influence of the width $L$ of the gas gap on real and
imaginary part of the dispersion relation $\lambda(k)$ for
$L=30,~31,~32,~\ldots,~40$. The current is $J_0=1.32\cdot 10^{-5}$.
All other parameters are as in Figs.~\ref{fig82}, \ref{fig83} and
\ref{fig84}: ${\cal R}_s=7000$, $C_s=0.243$, $L_s=54$, and ${\cal
U}_t=40$ at $L=36$.} \label{fig:disprel2}
\end{center}
\end{figure}

\subsubsection{Dependence on length $L$ of gas gap}

Fig.~\ref{fig:disprel2} shows for the same parameter values
that with increasing gas gap width $L$, the growth rate ${\rm
Re}\,\lambda(k)$ increases and the oscillation frequency
${\rm Im}\,\lambda(k)$ decreases, while the most unstable mode
stays nonvanishing: $k^*>0$. The explored gap lengths $L$ all
correspond to a falling current voltage characteristics of the gas discharge,
cf.~section~\ref{exp}.

\section{\label{sec:compar}Numerical solutions of the initial value problem}

\subsection{Implementation and results}

The full dynamical problem was also solved numerically
as an initial value problem. This allows us
to test the results of the stability analysis, to visualize the actual
dynamics and also
to study the behavior beyond the range of linear stability analysis.
Details of the numerical implementation are given in the appendix.

We study the case of high resistivity ${\cal R}_s=7000$
that leads to spatial pattern formation as discussed above.
Two values of the applied potential were investigated:
${\cal U}_{\rm t}=23.7$ where the homogeneous stationary state is stable,
and ${\cal U}_{t}=46.4$ where this state is Turing-Hopf-unstable.

In the transversal direction, we use periodic boundary conditions.
We choose the lateral extension $L_x$ as a multiple of the most unstable
wave length $2\pi/k^*$. After some tests with higher multiples
showing essentially the same behavior, we used $L_x=2\times 2\pi/k^*$,
i.e., the lateral extension is twice the expected wave length.
The initial condition is
\begin{equation}
\label{ic} \rho(x,z,0)=\rho_{0}(z)+C\,\rho_{1}(z)\,{\rm e}^{{\rm i}
k^*x}+{\rm c.c.},
\end{equation}
where $k^*$ is the wavenumber of the most unstable mode, $\rho_0(z)$ is the
stationary solution, and $\rho_1(z)$ is the eigenfunction for $k=k^*$
constructed in Sect.~\ref{subsec:ni}. The constant $C$ is chosen such that the
perturbation is small compared to $\rho_0(z)$. Note that we
need to specify the initial conditions for the ion density only,
since electron density and field are determined instantaneously
due to the adiabatic elimination of the fast electron dynamics.

Figure~\ref{fig:fig9} shows about one period of oscillation
within 4 time steps for the pattern forming case (${\cal U}_t=46.4$).
For each instant of time, the rescaled electron density $s=\sigma/\mu$
and the ion density $\rho$ are shown in the gas discharge region,
and the electric
field is shown both in the gas discharge and the semiconductor region,
as will be discussed in more detail later.
The upper row contains 3D plots and the lower row contour plots.
The figures show the characteristic electron and ion distribution
of a glow discharge, but with a strong spatio-temporal modulation.

The temporal period predicted by linear perturbation theory is 528.
This is agrees approximately with the numerical results.
On the other hand, the destabilization of the homogeneous stationary
state in Fig.~\ref{fig:fig9} is already far developed and in the fully
nonlinear regime. Therefore the results of the stability theory
at this time give only an indication for the full behavior.
In particular, the nonlinearity has created an onset to doubling
the spatial period that is absent for small perturbations.

\begin{figure}
\begin{center}
$\tau=7920$\\
\includegraphics[width=0.5\textwidth]{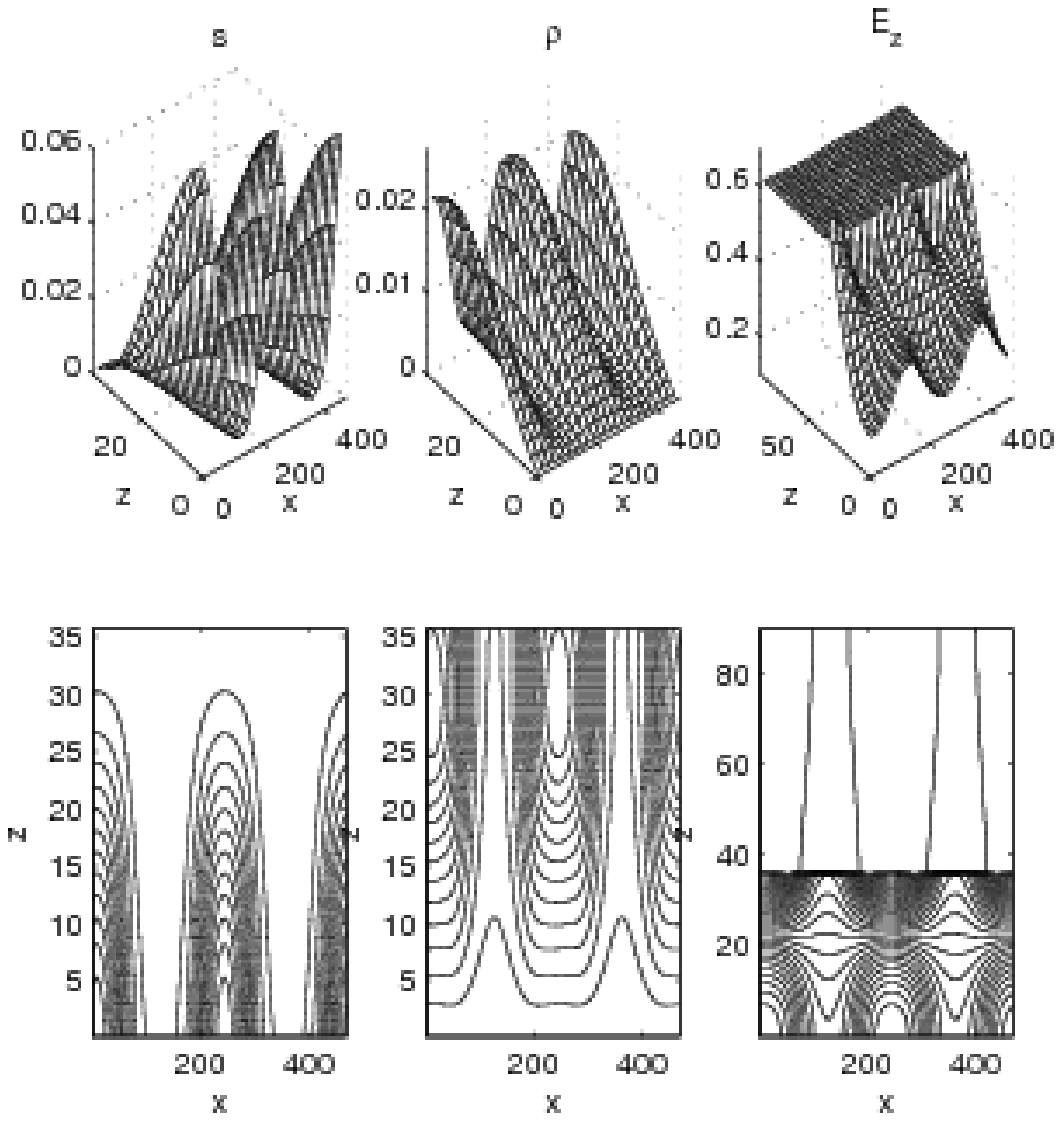} \\
$\tau=8160$\\
\includegraphics[width=0.5\textwidth]{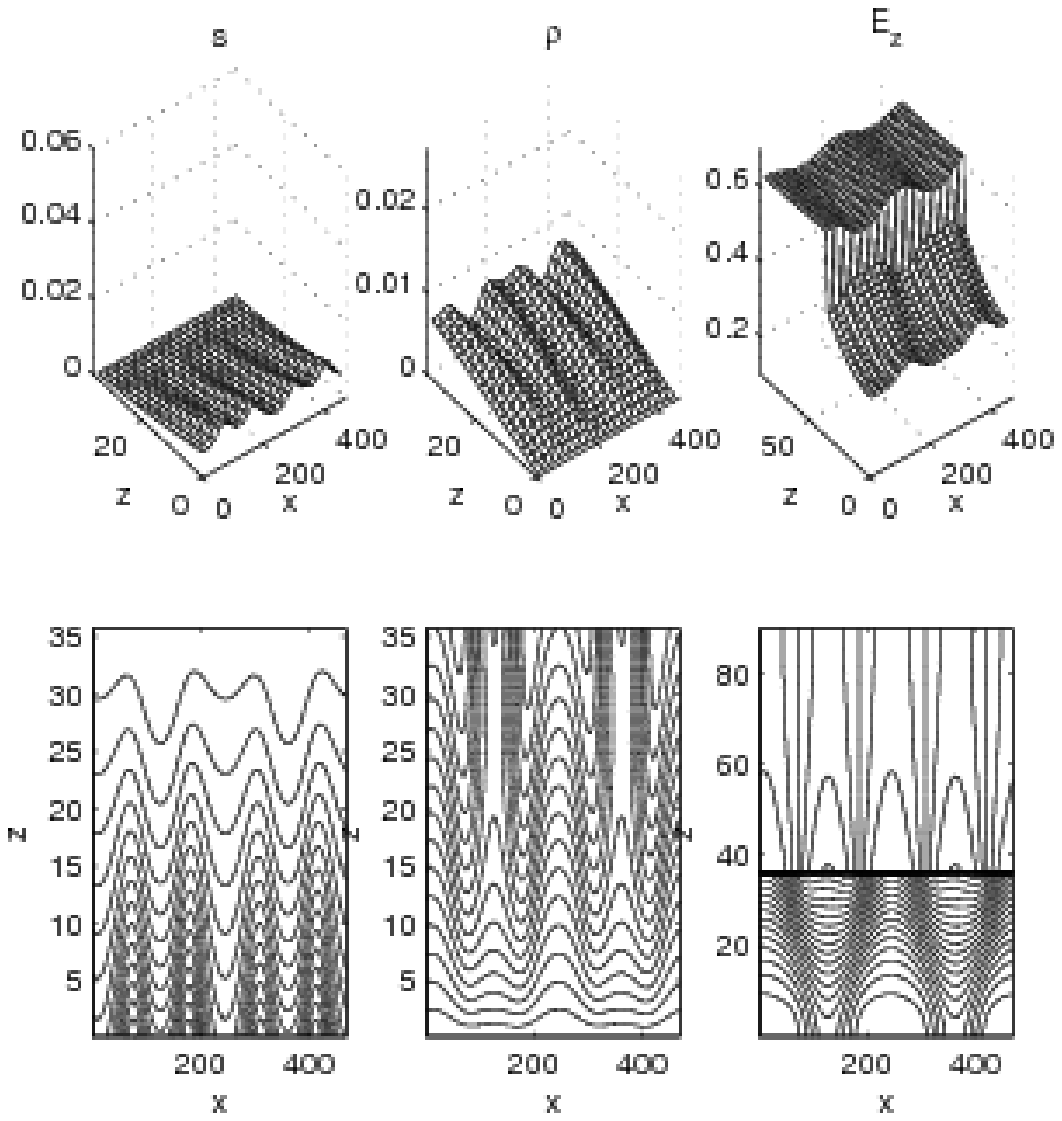}
\end{center}
\end{figure}
\begin{figure}
\begin{center}
$\tau=8040$\\
\includegraphics[width=0.5\textwidth]{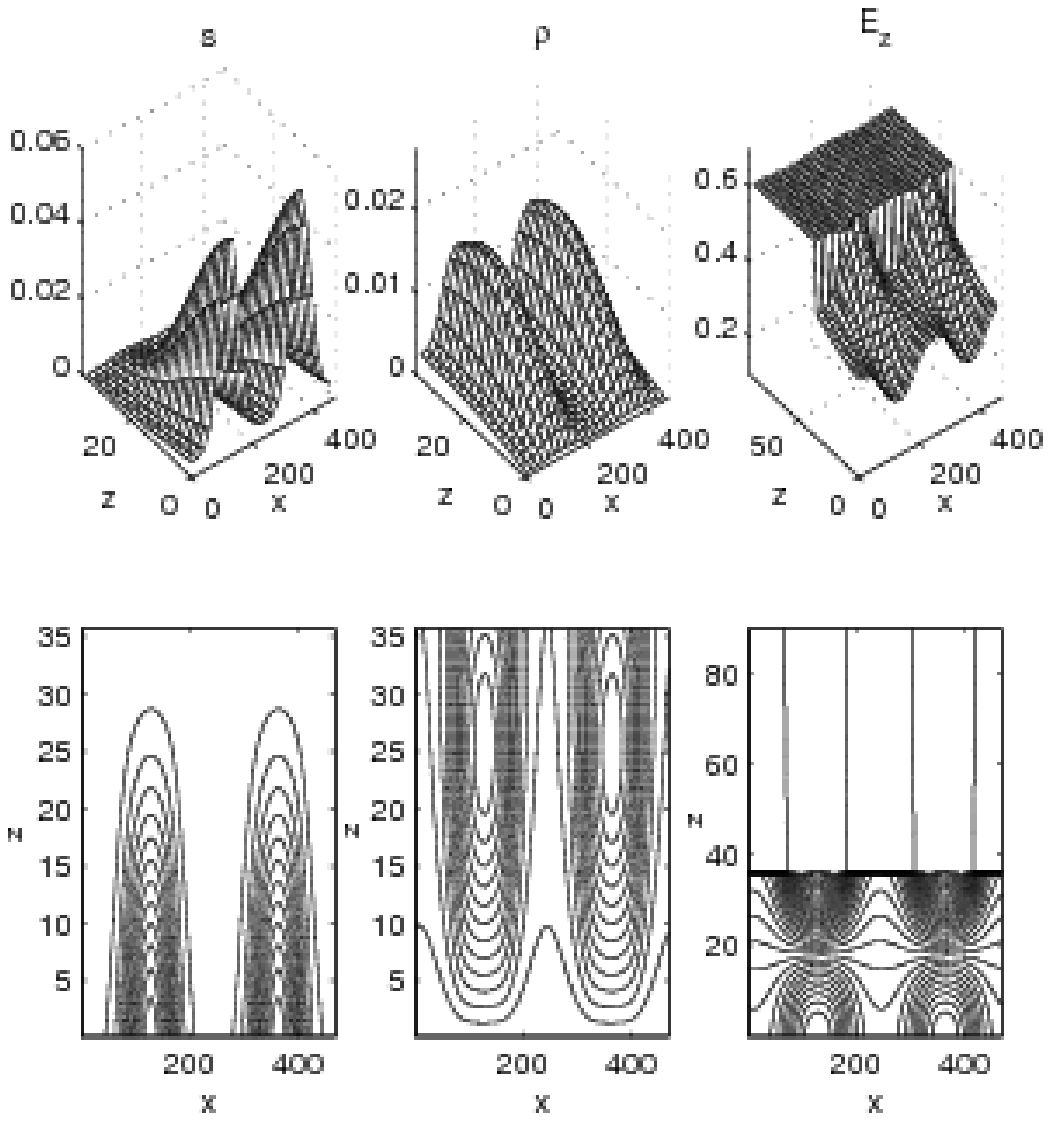} \\
$\tau=8280$\\
\includegraphics[width=0.5\textwidth]{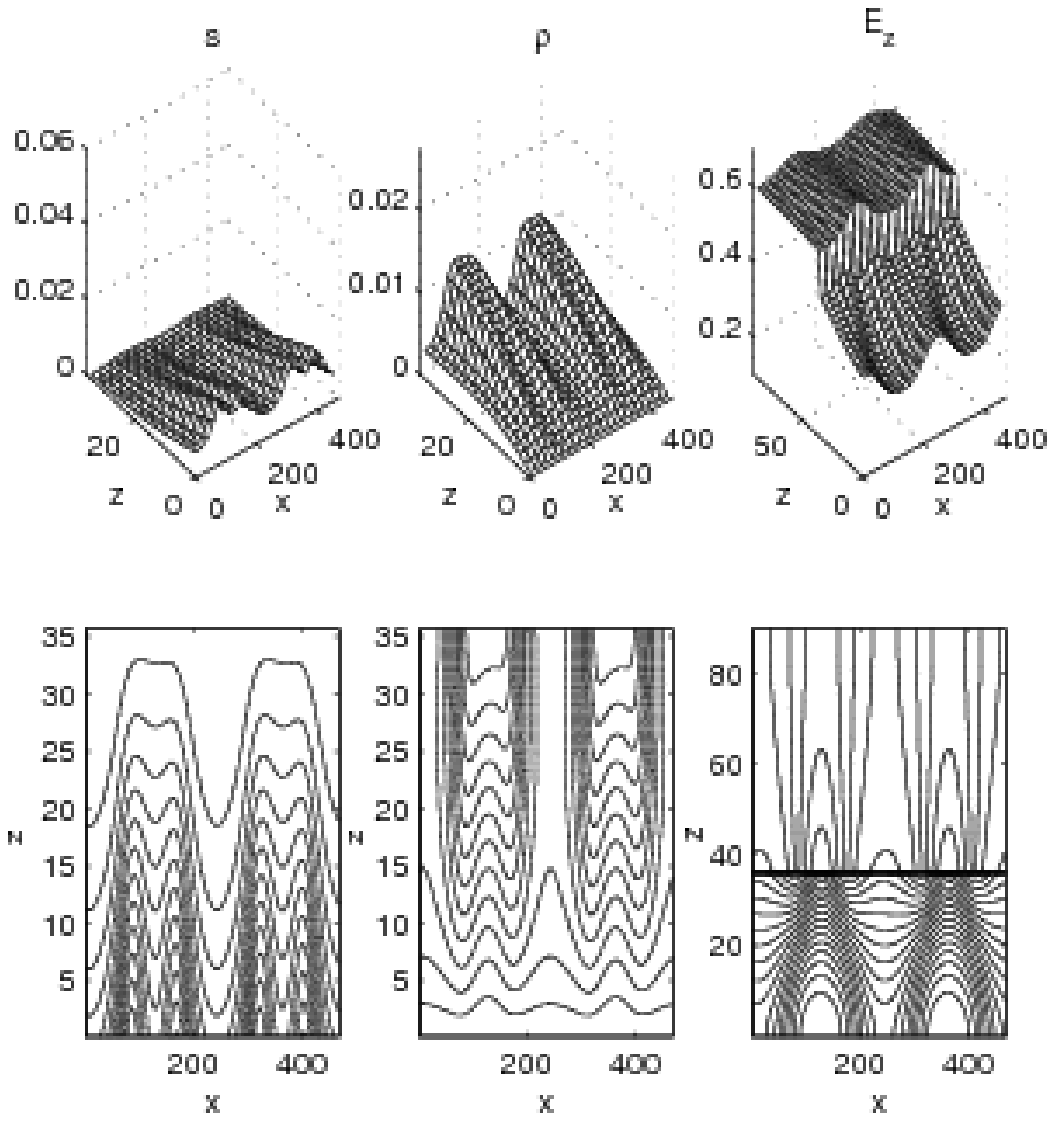} \\
\caption{ Profiles and contour lines of electron and ion particle
densities $s=\sigma/\mu$ and $\rho$ in the discharge region, and
electric field component $E_z$ in discharge and semiconductor region
at time steps $\tau=7920$, 8040, 8160, 8280 for ${\cal U}_{t}=46.4$
and ${\cal R}_s=7000$. $x$ and $z$ coordinates are as in Fig.~1 and
the text. For each time step, the data is represented as a 3D plot
in the upper row and as a contour plot in the lower row.}
\label{fig:fig9}
\end{center}
\end{figure}

\begin{figure}
\begin{center}
\includegraphics[width=0.5\textwidth
]{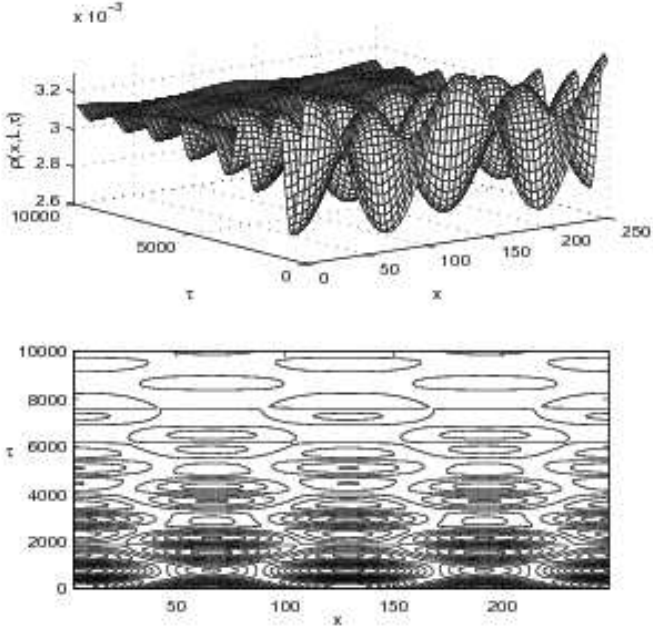} \\
\caption{Evolution of ion density $\rho(x,L,\tau)$ at the internal border $z=L$ for
${\cal U}_{t}=23.7$ and ${\cal R}_s=7000$ as a function of the
transversal coordinate $x$ and time $\tau$. Note that time $\tau$
increases towards the back.} \label{fig:fig6}
\end{center}
\end{figure}

\begin{figure}
\begin{center}
\includegraphics[width=0.5\textwidth
]{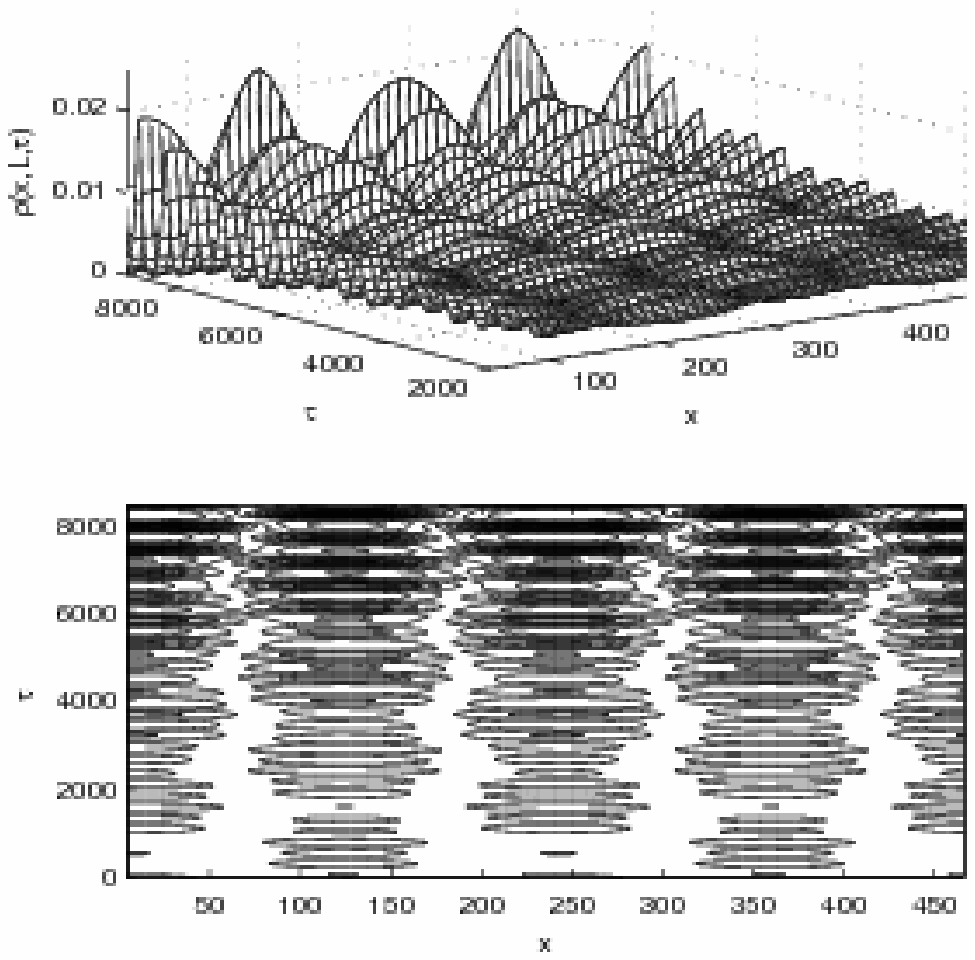} \\
\caption{The same as in the previous figure, but now for ${\cal
U}_{t}=46.4$; now the perturbations grow. Temporal snapshots
in the full $(x,z)$-plane of the
same numerical experiment are shown in Fig.~\ref{fig:fig9}.}
\label{fig:fig8}
\end{center}
\end{figure}

For presenting the evolution in time, the spatial structure has to be
represented on a line rather than in the full $(x,z)$-plane.
Obviously, the ion density on the semiconductor-gas-interface $(x,L)$
is an appropriate quantity, since it characterizes the local intensity
of the discharge glow in the transversal direction.
Figures~\ref{fig:fig6} and \ref{fig:fig8} show the complete temporal
evolution in such a presentation. Fig.~\ref{fig:fig6} presents data
of a perturbation decaying towards the stationary homogeneous state
for ${\cal U}_t=23.7$, while Fig.~\ref{fig:fig8} shows the growing
destabilization of the homogeneous stationary state for ${\cal U}_t=46.4$;
the late stage of this evolution was shown in Fig.~\ref{fig:fig9}.

\subsection{Comparison of numerical and stability results}

\begin{figure}
\begin{center}
\includegraphics[width=0.45\textwidth 
]{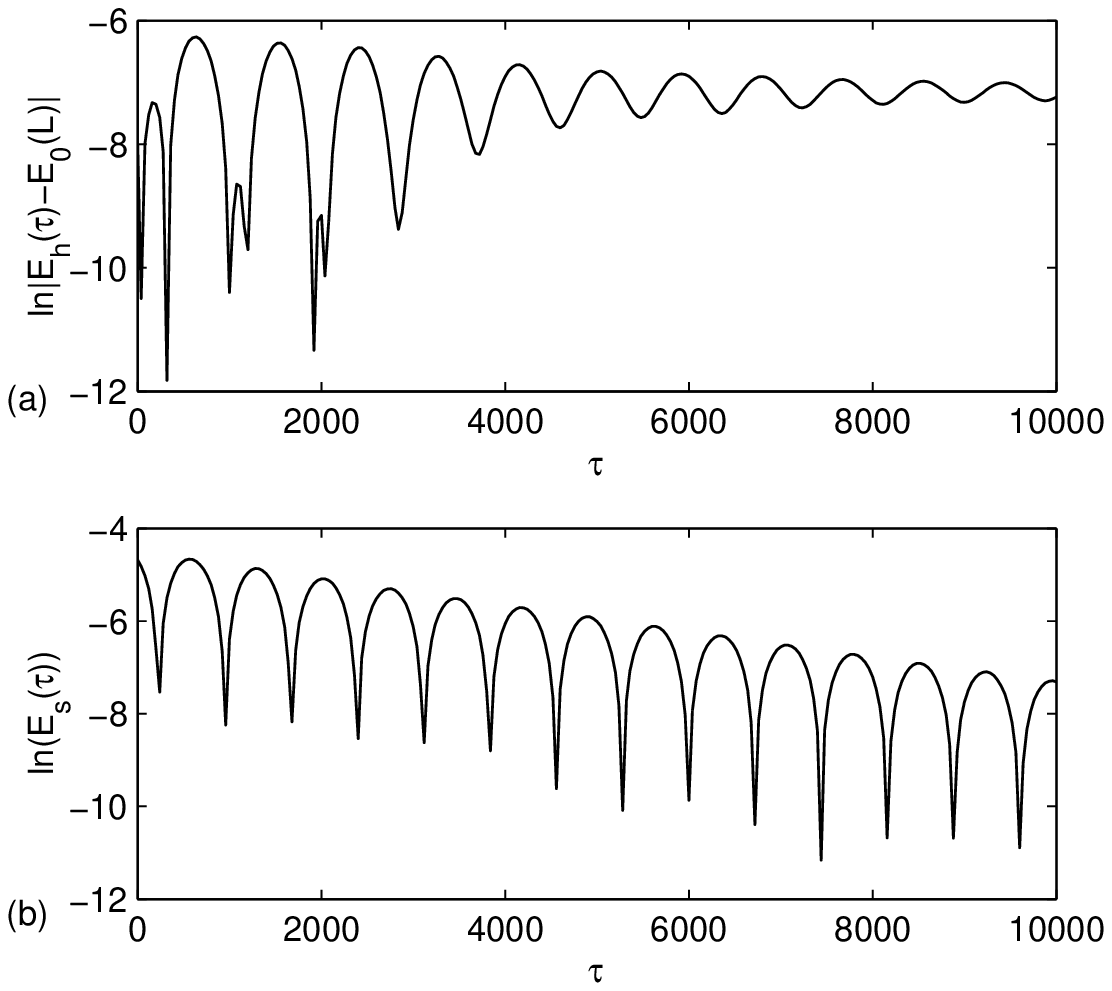} \\
\caption{Temporal evolution of the transversally averaged electric
field and of the spatial modulation of the field at the internal
border $z=L$ for ${\cal U}_{t}=23.7$ and ${\cal R}_s=7000$: (a)
$\ln |{\cal E}_{\rm h}(\tau)-{\cal E}_0(L)|$ and (b) $\ln {\cal
E}_{\rm s}(\tau)$ as a function of $\tau$.} \label{fig:fig4}
\end{center}
\end{figure}

\begin{figure}
\begin{center}
\includegraphics[width=0.45\textwidth 
]{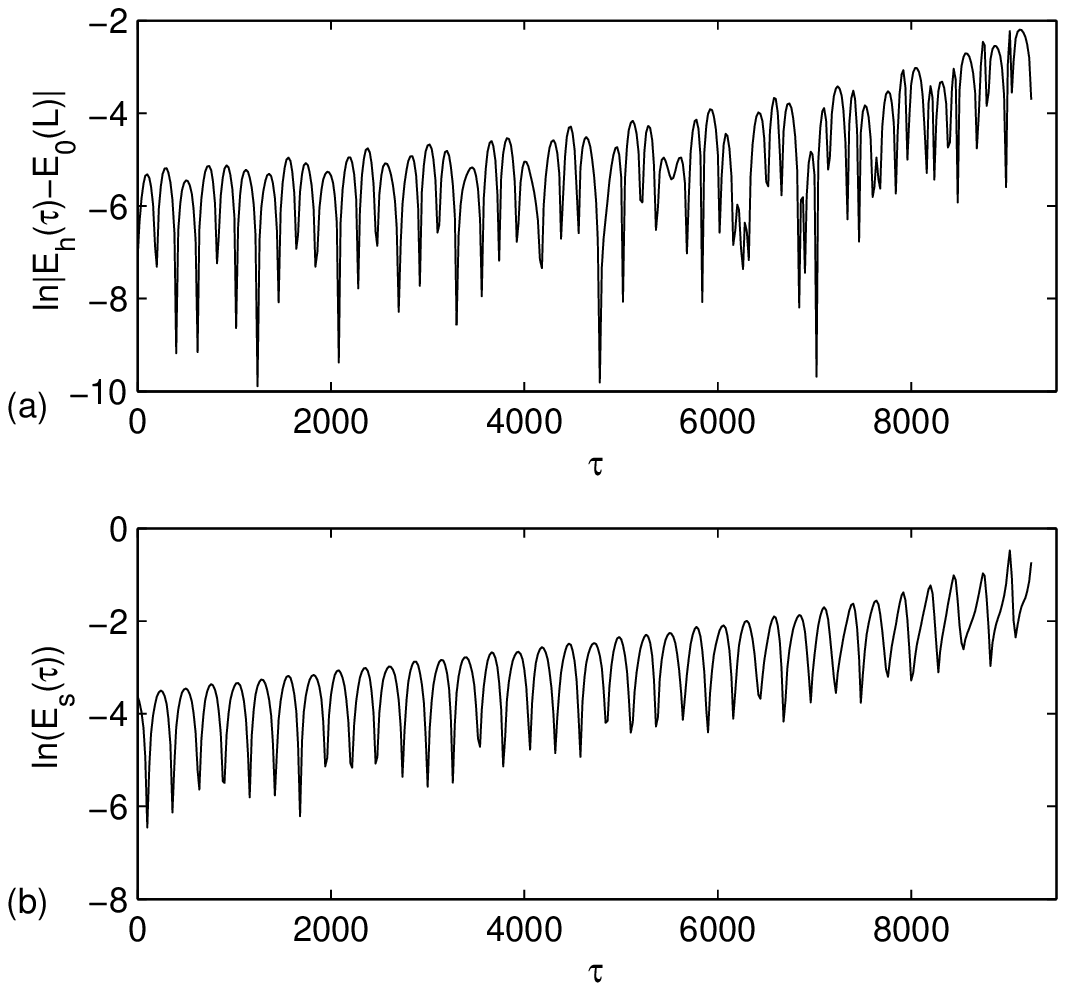} \\
\caption{The same as in the previous figure, now for ${\cal
U}_t=46.4$.} \label{fig:fig5}
\end{center}
\end{figure}

When one wants to compare results of the numerical simulation and
of the stability analysis, the evolution of different spatial modes
has to be extracted from the simulation. Appropriate quantities
are the transversally averaged electric field on the gas-semiconductor
interface
\begin{equation}
{\cal E}_{\rm h}(\tau)=\frac{1}{L_x}\int_0^{L_x}
{\cal E}_z\left(x,L,\tau\right)dx,
\end{equation}
and the spatial modulation of the field
\begin{equation}
{\cal E}_{\rm s}(\tau)=\max_x {\cal E}_z(x,L,\tau) -
\min_x {\cal E}_z(x,L,\tau).
\end{equation}
These quantities, or rather the logarithms of $|{\cal E}_{\rm
h}(\tau)-{\cal E}_0(L)|$ and of ${\cal E}_s(\tau)$ are shown for the
stabilizing case ${\cal U}_t=23.7$ in Fig.~\ref{fig:fig4}, and for
the destabilizing case
${\cal U}_t=46.4$ in Fig.~\ref{fig:fig5}.

In this logarithmic plot for the fields, the lines through
the maxima are approximately straight, which means
that the growth is exponential. For the destabilizing case,
the growth rate of the spatial mode ${\cal E}_{\rm s}(\tau)$
is slightly larger than that of the homogeneous mode
${\cal E}_{\rm h}(\tau)$  which implies that the most unstable
mode has a non-vanishing wave number $k^*$:
${\rm Re}\,\lambda(k^*) > {\rm Re}\,\lambda(0)$ in agreement with
linear stability analysis.
Furthermore, at late stages when the dynamics is beyond the range
of linear perturbation theory and becomes nonlinear, the growth
of all modes accelerates.

\begin{figure}
\begin{center}
\includegraphics[width=0.5\textwidth
]{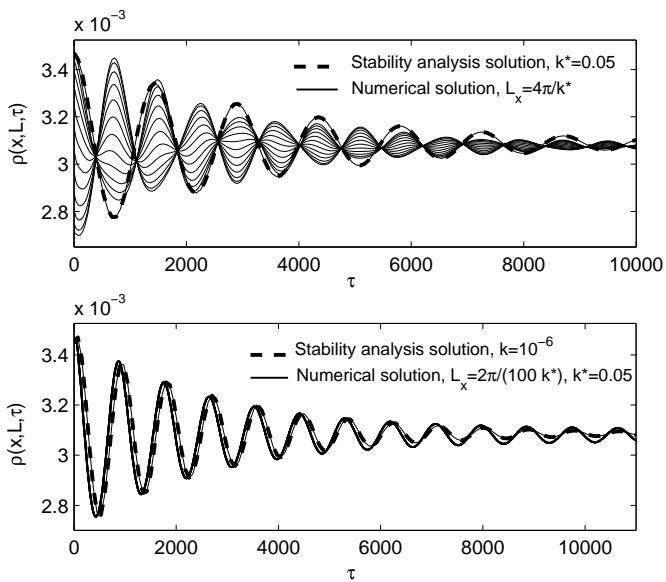} \\
\caption{(Color online) Comparison of results of the PDE solutions
(solid lines) and of the stability analysis (dashed line). Ion
density $\rho$ at the computational nodes between $x=0$ and
$x=L_x/4$ of the internal border $z=L$ as a function of time for
${\cal U}_{t}=23.7$ and ${\cal R}_s=7000$.} \label{fig:fig2}
\end{center}
\end{figure}

\begin{figure}
\begin{center}
\includegraphics[width=0.5\textwidth
]{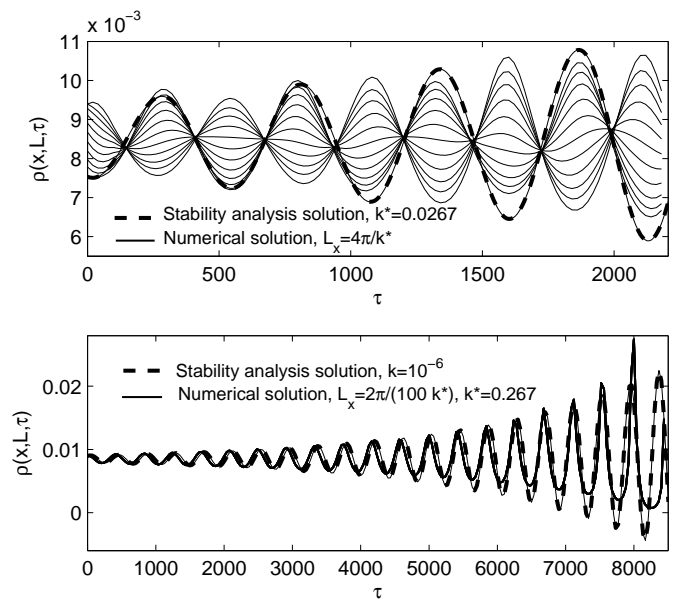} \\
\caption{(Color online) The same as in the previous figure, now for
${\cal U}_{t}=46.4$.} \label{fig:fig3}
\end{center}
\end{figure}

Figs.~\ref{fig:fig2} and \ref{fig:fig3} show a quantitative
comparison between stability analysis and computational results.
Fig.~\ref{fig:fig2} shows the stabilizing case ${\cal U}_{t}=23.7$.
The stability analysis predicts that $k^*=0.050$ is the most
unstable mode; it has the eigenvalue $\lambda(k^*)=-0.2807\cdot
10^{-3}+0.4320\cdot 10^{-2}\,{\rm i}$. Therefore, the period of the
temporal oscillations is predicted as $2\pi/{\rm Im}(\lambda)=
1454$, the characteristic decay time as $1/{\rm Re}(\lambda) =
3563$, and the characteristic wave length as $2\pi/k^*= 126$. This
predicted behavior is shown as the dashed line in the upper panel of
Fig.~\ref{fig:fig2}. The solid lines show the numerical solution,
more precisely the time evolution of the ion density
$\rho(x,L,\tau)$ evaluated on the grid nodes in the range between
$x=0$ and $x=L_x/4$ on the gas-semiconductor interface $z=L$. Period
and growth rate agree quantitatively, therefore both simulations and
stability analysis can be trusted.

The predictions on the $k{=}0$-mode are tested in the lower panel
of Fig.~\ref{fig:fig2}: here the transversal extension of the simulation
system was chosen so narrow that transversal modes had no space
to develop: the width was taken as $L_x=2\pi/(100 k^*)$ where $k^*$
is the most unstable wave length. In this case, only the $k=0$-mode
can grow, it has $\lambda(0)=-0.3547\cdot 10^{-3}+0.7102\cdot
10^{-2}\,{\rm i}$. The plot again shows a very good agreement
between stability analysis and simulation, now effectively for
the one-dimensional case.

Finally, in Fig.~\ref{fig:fig3} again the destabilizing state
for ${\cal U}_{t}= 46.4$ is shown. The stability analysis predicts
the most unstable wave number $k^*=0.0267$ and its eigenvalue
$\lambda(k^*)=0.4615\cdot 10^{-3}+0.1191\cdot 10^{-1}\,{\rm i}$.
The two panels show again the predicted and the simulated
oscillations in a laterally wide system allowing the formation
of the $k^*$-mode, and in the narrow system that only has space
for the $k{=}0$-mode. Again, the agreement is convincing.


%
%
%
%

\section{\label{sec:concl} Conclusions and discussions}

We have investigated the onset or decay of spatio-temporal patterns
in a layered semiconductor-gas discharge system subject to a DC voltage.
By means of linear stability analysis, we were able to derive complete
phase transition diagrams, rather than only to investigate single points
in parameter space in a simulation.

We have used the simplest model possible:
Only the drift motion of electron and ion densities
is taken into account, and the electrons are adiabatically eliminated.
The semiconductor is approximated as a linear Ohmic conductor,
and nonlinear effects come in only through the space charges
of the ions in the gas discharge gap and through the surface charges
on the gas-semiconductor interface. (We remark that particle diffusion
has a smoothening effect and is not expected to generate any new
structures. However, effects like gas heating or nonlinear semiconductor
characteristics can create additional destabilization mechanisms.)

Methods and results of the linear stability analysis of the homogeneous stationary
state and of the full numerical simulation of the initial value problem
are presented. The choice of parameters is guided by the experiment
described in \cite{Str2}; they are summarized in Sect.~\ref{param}.
In the experiment \cite{Str2}, the resistance of the semiconductor
can be changed by a factor of 10 by photo-illumination without
changing any other system parameter, and a full phase transition diagram is derived
experimentally. It is seen that the system never
relaxes to a spatially structured time-independent state, but depending
on the resistance, it either forms a homogeneous oscillating or
a spatially structured oscillating state.

Like in the experiments, the homogeneous stationary state is either
stable, or it can be destabilized by a temporal (Hopf) or spatio-temporal
(Turing-Hopf) mode; a purely spatial destabilization (Turing) is observed
neither in experiments nor in theory. The transition from stable to unstable
behavior is in about the same parameter regime in a diagram spanned by
applied voltage ${\cal U}_t$ and inverse resistance of the semiconductor
$1/{\cal R}_s$. However, the parameter range where either Hopf- or
Turing-Hopf-destabilization is predicted, does not agree with the experimental
range where purely temporal or spatio-temporal patterns are observed.
This is discussed in detail in Section~\ref{CompExp}. A possible explanation
is that we might be comparing the wrong transitions. The theoretical prediction
of linear perturbation theory concerns a destabilization to weak running waves
of long wave lengths. These waves resemble much more the "diffuse" moving bands
reported only in the Ph.D.~thesis \cite{StrTh}, than the very nonlinear small
dynamic filament structures described in \cite{Str2}. This suggestion actually
asks for further experimental investigations.
It is interesting to note that the physical mechanism of
these ``diffuse'' bands has nothing to do with particle diffusion,
but only with the Laplacian nature of the created electric fields.
For further predictions on the parameter dependence of the linear
instabilities, we refer to Section \ref{ParamDisp}.

Our numerical solutions of the initial value problem in Section~\ref{sec:compar} agree well with our linear
stability analysis within its range of validity. First of all, this proves the correct implementation of
both methods. Second, for larger amplitudes, new nonlinear spatial structures appear such as the spatial
period doubling in Fig.~\ref{fig:fig9}. For ${\cal R}_s=7000$ and ${\cal U}_t=40$, these oscillations
actually have been seen to reach a limit cycle that corresponds to a standing wave. Of course,
the full nonlinear behavior in three spatial dimensions should be investigated in the future,
but the 2D results give a first indication for the behavior.

We remark finally that only the simplest possible model with
nonlinear space charge effects was investigated. Of course,
the model can be extended by various additional mechanisms,
but obviously the simple model already contains all relevant physics
to predict the onset of pattern formation in the correct parameter regime.

\begin{acknowledgments}
The authors thank Alexander Morozov from Instituut-Lorentz, Leiden University,
for advice on the numerical method described in Section~\ref{strat} and for
help with its implementation. I.R. acknowledges numerous discussions with
W. Hundsdorfer at the Center for Mathematics and Computer Science (CWI)
Amsterdam on numerical mathematics. I.R. acknowledges financial support by
ERCIM and FOM, and D.S. by FOM for their work in Amsterdam where the
numerical stability analysis was performed and the numerical PDE
work was initiated.
\end{acknowledgments}

\appendix

\section{\label{sec:num}Numerical procedure}

Here we describe the numerical method used for solving
the initial value problem numerically.
The computation is based on a finite-difference technique
to solve equations (\ref{resc2})--(\ref{resc23}) with boundary conditions
 (\ref{2dg05})--(\ref{jump1}) and periodic boundary conditions in the
transversal direction.

\begin{figure}
\begin{center}
\includegraphics[width=0.45\textwidth]{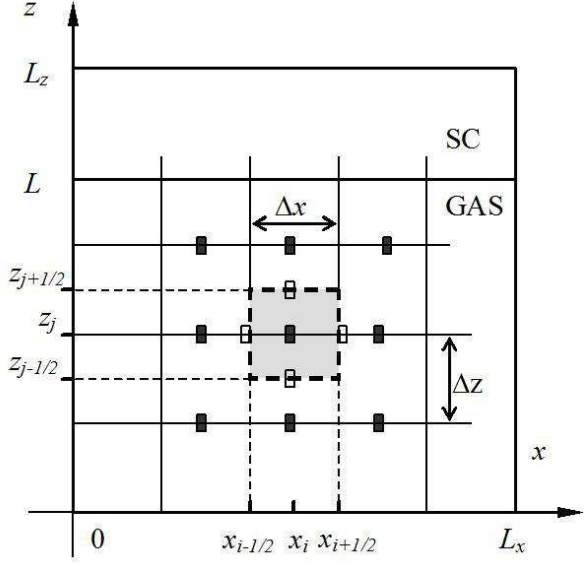} \\
\caption{Computational domain and computational cell.}
\label{fig:fig1}
\end{center}
\end{figure}

The computational domain is a rectangular region $[0,L_x] \times
[0,L_z]$ on a two-dimensional Cartesian coordinate system $(x,z)$,
which consists of two layers -- gas and semiconductor, see Figs.~\ref{fig1}
and \ref{fig:fig1}. We use a uniform vertex-centered grid in the
'vertical' $z$-direction with nodes
$$ z_j = j \Delta z,~~~ \Delta z=\frac{L_z} N,~~~j=0,1, \cdots,N~$$
and a uniform cell-centered grid with nodes
$$ x_i= \left(i - \frac{1}{2}\right) \Delta x ,~~~ \Delta x= \frac{L_x}M,~~~i=1,2, \cdots,M~$$
on the 'horizontal' x-direction. The grid is spaced such that the
internal interface between semiconductor and gas region lies exactly on
the grid line.

The densities $s=\sigma/\mu$ and $\rho$ and the electric
potential $\phi$ are evaluated on the nodes
of the grid, while the $x$ and $z$ components of the electric field
(${\cal E}_x$ and ${\cal E}_z$, respectively) are calculated on
the surfaces of the computational cell (Fig.~\ref{fig:fig1}).

To obtain a finite-difference representation of equations
(\ref{resc2}) and (\ref{resc22}), we first integrate them over the
cell volume $ x_{i-1/2} \leq x \leq  x_{i+1/2},~ z_{j-1/2} \leq z
\leq z_{j+1/2}$. Let us consider in detail the equation for
the ions (\ref{resc22}). After its integration, we come to
\begin{eqnarray}
\frac{d \rho_{j,i}}{d \tau} &=& \frac{({\cal E}_{\rm
x}~\rho)_{j,i-1/2} - ({\cal E}_x~\rho)_{j,i+1/2}}{\Delta x}
\nn\\
&&+\frac{({\cal E}_z~\rho)_{j-1/2,i} - ({\cal E}_{\rm
z}~\rho)_{j+1/2,i}}{\Delta z} + f_{j,i}~.
\end{eqnarray}
The subscripts $i$ and $j$ are related to $x$ (transversal) and $z$
(longitudinal) directions respectively and $f$ stands for the source
term of Eq. (\ref{resc22}).

The choice of $\rho_{j \pm 1/2,i}$ and $\rho_{j,i\pm 1/2}$ at the
surfaces of the computational cell determines the concrete
discretization method for the convective terms of Eq.
(\ref{resc22}). We used the third-order upwind-biased scheme (see,
e.g., \cite{Willem}, p.~83), which in $z$- and $x$-direction is given by
\begin{eqnarray}
\label{convz}
\left({\cal E}_z\rho\right)_{j+1/2,i} &=&
\frac{1}{6} \Big[{\cal E}^{+}_{z~ j+1/2,i} \left(-\rho_{j-1,i} + 5
\rho_{j,i} +2 \rho_{j+1,i}\right) \nn \\
&&+ {\cal E}^{-}_{z~j+1/2,i} \left(2\rho_{j,i}
+ 5 \rho_{j+1,i} - \rho_{j+2,i}\right)\Big],
\nn\\
\label{convx}
\left({\cal E}_x\rho\right)_{j,i+1/2} &=&
\frac{1}{6} \Big[{\cal E}^{+}_{x~ j+1/2,i} \left(-\rho_{j,i-1} + 5
\rho_{j,i} +2 \rho_{j,i+1}\right) \nn \\
&& + {\cal E}^{-}_{x~
j,i+1/2} \left(2\rho_{j,i} + 5 \rho_{j,i+1} - \rho_{j,i+2}\right)\Big].
\nn\\ &&
\end{eqnarray}
Here, the electric field components are
$$
{\cal E}^+_{\ldots} = \max\Big[~0,~{\cal E}_{\ldots}~\Big], ~~~
{\cal E}^-_{\ldots} = \min\Big[~0,~{\cal E}_{\ldots}~\Big],
$$
and $\bs{\cal E}=-\nabla\phi$ is discretized as
\begin{eqnarray}
{\cal E}_{z\,j+1/2,i} &=&  - \:\frac{\phi_{j+1,i}-\phi_{j,i}}{\Delta z},
\nn\\
{\cal E}_{x\,j,i+1/2} &=& -\:
\frac{\phi_{j,i+1}-\phi_{j,i}}{\Delta x}.
\end{eqnarray}

For the numerical time integration, we used the extrapolated second
order BDF2 method, see \cite{Willem}, p.~204, \cite{Wes}, p.~197,
whose variable step size version has the form
\begin{eqnarray}
\label{time}
&&\rho^{m}-\frac{\left(1+r\right)^{2}}{1+2r}\rho^{m-1} +
\frac{r^2}{1+2r}\rho^{m-2} \nonumber \\
&&=\frac{\left(1+r\right)}{1+2r} \Delta \tau_{m} ~ (2F^{m-1}
-F^{m-2})~,~~m \geq 2,
\end{eqnarray}
where the superscript $m$ denotes the time $\tau_m$ with step size
$\Delta \tau_m=\tau_m-\tau_{m-1}$ and step size ratio
$r=\Delta\tau_m/\Delta\tau_{m-1}$. Here $F$ contains the discretized
convective terms and a source term. Note that we have dropped
spatial indices in Eq.~(\ref{time}). Since the two-step method needs
$\rho^0$ and $\rho^1$ as starting values, the explicit Euler method
\be \label{euler} \rho^{m} = \rho^{m-1} + \Delta \tau_m ~
F(\tau_{m-1}, \rho^{m-1}) \ee is used for the first step $m=1$.
Because of the explicit time integration, we are restricted by the
standard CFL stability condition.

The same space discretization technique and time integration
method are used also for the electron density equation
(\ref{resc2}) that contains no temporal derivative. Note that
in this case the $z$-direction plays
the role of 'time'  in (\ref{time}) and (\ref{euler}).

To obtain a finite-difference approximation for Poisson's equation
(\ref{resc23}), we use the traditional second order discretization:
\begin{eqnarray}
&&-\frac{\phi^m_{j,i-1}-2\phi^m_{j,i}+ \phi^m_{j,i+1} }{(\Delta
x)^2}\; - \frac{\phi^m_{j-1,i}-2\phi^m_{j,i}+ \phi^m_{j+1,i}
}{(\Delta z)^2} \nn  \\
\label{eq:Poisson}
&&\qquad\qquad =\left\{
\begin{array}{ll}
0~, & \textrm{gas-discharge layer,}  \\
\rho^{m-1}_{j,i} ~, & \textrm{semiconductor layer.}
\end{array} \right.
\end{eqnarray}

This equation is valid everywhere except at the gas semiconductor
interface where one has to account for a finite surface charge
as well as for a discontinuity of the dielectricity constant.
On this interface, the discrete version of the 'jump' condition
(\ref{jump}) is used instead of (\ref{eq:Poisson}).
The system of resulting difference equations is solved by a
symmetrical successive over-relaxation method (SSOR), see \cite{Thomas},
p.~343.

The complete numerical procedure was organized as follows. For every
new $(m+1)$th time step, first Poisson's equation was solved using
the known ion density $\rho^{m}$ and surface charge value
$q^{m}_{b}$ in the jump condition (\ref{jump}), determining the
electric field components in the new time step. Second, the electron
density in the new time step $s^{m+1}$ was calculated. This
determined the source term in the continuity equation for ions.
Third the ion density $\rho^{m+1}$ was calculated, which finally
determined the new value for the surface charge in (\ref{jump}).

The numerical convergence was checked by performing several
calculations with different error tolerance parameters for Poisson's
equation, using refinement of the space grid, and different time
stepping parameters. The number of grid nodes used in the
calculations was $52\times 361$ in the $x$ and $z$ directions,
respectively, for the potential in the whole gas discharge -
semiconductor region, and $54\times 147$ in the $x$ and $z$
directions for the particle densities in the gas discharge region.
When solving Poisson's equation, the iteration process is stopped
when the relative error is
$\|\phi^{(k+1)}-\phi^{(k)}\|/\|\phi^{(k+1)}\|< 5\cdot 10^{-7}$.


\end{document}